%% file: main.tex
\documentclass[letterpaper,hmargin=1.05in,vmargin=1.05in]{article}

\usepackage[margin=1in]{geometry}
\usepackage[T1]{fontenc}
\usepackage{lmodern}
\usepackage{microtype}
\usepackage{amsmath,amssymb,amsthm,mathtools}
\usepackage{booktabs}
\usepackage{enumitem}
\usepackage{xcolor}
\usepackage{comment}
\usepackage[hidelinks]{hyperref}
\usepackage[capitalise,noabbrev]{cleveref}
\hypersetup{
 pdftitle={Classical Key Agreement with Near-Quadratic Quantum Security},
 pdfauthor={Samuel Bouaziz{-}{-}Ermann},
 pdfsubject={Classical Key Agreement with Near-Quadratic Quantum Security}
}
\usepackage{authblk}

\newtheorem{theorem}{Theorem}[section]
\newtheorem{lemma}[theorem]{Lemma}
\newtheorem{proposition}[theorem]{Proposition}

\newtheorem{remark}[theorem]{Remark}

\input{macro2}

\title{Classical Key Agreement with Near-Quadratic Quantum Security}
\author{Samuel Bouaziz{-}{-}Ermann}
\affil{Okinawa Institute of Science and Technology Graduate University, Japan
\authorcr{\small samuel.bouazizermann@oist.jp}}
\date{}

\begin{document}
\maketitle

\begin{abstract}
  In 1974, Merkle proposed a key exchange protocol where the honest parties make \(N\) queries to a random oracle to compute a secret key, while the best attacker needs to make \(\Omega(N^2)\) queries to recover the key with constant probability.
  This protocol was later proved to be optimal by Barak and Mahmoody (Crypto'09).
  In the quantum setting, Belovs et al. (TQC'17) proved that for any \(\varepsilon >0\), there exists a classical protocol with \(O(N)\) expected honest queries, for which any quantum eavesdropper making \(O(N^{1.5-\varepsilon})\) queries has vanishing key recovery probability.

  In this paper, we give two families of classical key-establishment protocols in the
random oracle query model.
The first is a two-message protocol and the second is a three-message protocol with
perfect agreement.
In both cases the
honest parties make $\widetilde O(\securityParam )$ classical oracle queries, whereas every
constant-success quantum eavesdropper needs \(\widetilde\Omega(\securityParam ^2)\)
quantum queries to recover the  key.
Our result is tight up to logarithmic factors, as the quadratic classical attack of Barak and Mahmoody applies in this setting.

The construction combines two levels of hidden sum relations.
In the two-message protocol, Alice samples random labels and encodes
their bits using masked oracle blocks.
A zero is encoded by a uniformly random
block, while a one is encoded by a block containing a planted sum relation.
Bob decodes a random subset of the labels and publishes their sum.
Alice uses her original labels to identify this subset, which becomes
their shared key.

We also give an efficient implementation with $O(N\log N)$ honest
oracle queries and $O(N(\log N)^2)$ combined honest running time,
for which recovering the key with constant probability requires
$\Omega(N^{4/3})$ quantum queries.
\end{abstract}

\clearpage
\setcounter{tocdepth}{2}
\tableofcontents
\clearpage

\input{tex/intro2}

\input{tex/preliminaries2}

\input{tex/proof2}

\input{tex/perfect2}

\input{tex/near-quadratic2}

\bibliographystyle{alpha}
\bibliography{./cryptobib/abbrev3,./cryptobib/crypto,references}

\end{document}

%% file: macro2.tex
\newcommand{\F}{\mathbb F}

\newcommand{\Prb}{\mathop{\rm Pr}}

\newcommand{\supp}{\operatorname{supp}}
\newcommand{\wt}{\operatorname{wt}}

\newcommand{\Unif}{\operatorname{Unif}}

\newcommand{\cP}{\mathcal P}
\newcommand{\cQ}{\mathcal Q}

\newcommand{\cX}{\mathcal X}
\newcommand{\cY}{\mathcal Y}

\newcommand{\eps}{\varepsilon}
\newcommand{\ket}[1]{\lvert #1\rangle}

\newcommand{\securityParam}{N}
\newcommand{\sumSize}{m}
\newcommand{\sumArity}{k}
\newcommand{\outerSize}{n}
\newcommand{\outerArity}{\ell}
\newcommand{\innerSize}{M}
\newcommand{\innerArity}{r}
\newcommand{\outerBits}{q}
\newcommand{\innerBits}{s}
\newcommand{\numBlocks}{d}

\newcommand{\plantedDist}{P}
\newcommand{\nullDist}{Q}
\newcommand{\plantedSpace}{\cP}
\newcommand{\nullSpace}{\cQ}
\newcommand{\sqrtDist}{\delta}
\newcommand{\bitDist}{\mu}
\newcommand{\acceptProb}{s}
\newcommand{\outputOverlap}{\tau}
\newcommand{\successProb}{\nu}

\newcommand{\advMatrix}{\Gamma}
\newcommand{\queryDiff}{\Delta}
\newcommand{\sumCoeff}{\alpha}
\newcommand{\sumScale}{L}
\newcommand{\sumBlock}{G}
\newcommand{\fourierVector}{e}
\newcommand{\restrictedVector}{\phi}
\newcommand{\character}{\chi}
\newcommand{\fourierProjector}{E}
\newcommand{\sumGroup}{G}
\newcommand{\groupSize}{g}
\newcommand{\gramMatrix}{H}
\newcommand{\innerAdv}{A}
\newcommand{\outerAdv}{B}
\newcommand{\composedAdv}{C}
\newcommand{\innerNorm}{\sigma}

\newcommand{\impProtocol}{\Pi^{\rm imp}}
\newcommand{\perfProtocol}{\Pi^{\rm perf}}
\newcommand{\randomOracle}{H}
\newcommand{\publicMask}{m}
\newcommand{\sharedKey}{I}
\newcommand{\queryBudget}{T}
\newcommand{\honestBudget}{H}
\newcommand{\relationTest}{\operatorname{Rel}}
\newcommand{\bitDecoder}{\operatorname{Dec}}
\newcommand{\pairDecoder}{\operatorname{PairDec}}

\providecommand{\bra}[1]{\langle #1\rvert}
\providecommand{\braket}[2]{\langle #1\mid #2\rangle}
\providecommand{\ketbra}[2]{\ket{#1}\!\bra{#2}}
\newcommand{\matrixel}[3]{\bra{#1}#2\ket{#3}}
\newcommand{\adj}{\dagger}
\newcommand{\identity}{\mathbb I}
\newcommand{\labelSpace}[1]{\mathcal H_{#1}}
\newcommand{\restrictionOp}{R}
\newcommand{\fourierSupport}{T}

\newcommand{\distKet}[1]{\ket{\sqrtDist_{#1}}}
\newcommand{\distBra}[1]{\bra{\sqrtDist_{#1}}}
\newcommand{\fourierKet}[1]{\ket{\fourierVector_{#1}}}
\newcommand{\fourierBra}[1]{\bra{\fourierVector_{#1}}}
\newcommand{\restrictedKet}[2]{\ket{\restrictedVector_{#1,#2}}}

\newcommand{\fallbackEvent}{\mathcal F}

%% file: tex/intro2.tex
\providecommand{\fallbackKey}{0}
\providecommand{\agreedKey}{K}
\providecommand{\blockTarget}{w}
\providecommand{\recoveryScale}{\mathcal L}

\section{Introduction}
\label{sec:introduction}

In the classical setting, Merkle puzzles are a key exchange protocol that gives a quadratic gap between the work
of two honest classical parties and that of a classical
eavesdropper~\cite{Merkle78}.
In the random oracle setting, the honest parties make
\(O(\securityParam)\) queries to a random oracle to compute a secret key, while a
classical eavesdropper needs $\Omega(\securityParam^2)$ queries to
recover the key with constant probability.
Here, \(\securityParam\) is the security parameter.

The quadratic gap is optimal against classical eavesdroppers.
Impagliazzo and Rudich established the basic barrier to constructing classical
key agreement from one-way functions by black-box methods~\cite{STOC:ImpRud89}.
Barak and Mahmoody later proved the optimal quantitative
bound in the random oracle model~\cite{C:BarMah09}.
If each honest party makes at most $\securityParam$ queries and their
agreement probability is $\rho$, then, for every fixed $\delta>0$,
a classical eavesdropper can recover the key with probability
at least $\rho-\delta$ using $O_\delta(\securityParam^2)$ queries.
This bound applies to protocols with any number of
communication rounds.

An
eavesdropper that can query the oracle in superposition can use Grover's search to recover the key in the original Merkle construction using only \(O(\securityParam)\) quantum queries~\cite{Grover,BrassardSalvail08}.
Brassard and Salvail showed that a superlinear advantage can be
restored if the honest parties can also use quantum
computation~\cite{BrassardSalvail08}.
Their protocol uses $O(\securityParam)$ honest queries and requires
$\Omega(\securityParam^{3/2})$ quantum queries for key recovery
with constant probability.
Brassard, H\o yer, Kalach, Kaplan, Laplante, and Salvail then developed further
Merkle-style protocols, including protocols with entirely classical honest parties and superlinear resistance to quantum attacks~\cite{C:BHKKLS11,JC:BHKKLS19}.
A central difficulty is proving security for the instances that
the protocol actually produces.
A worst-case query lower bound only guarantees that some inputs
are hard.
It does not show that an eavesdropper usually fails in
a random execution.
Belovs, Brassard, H\o yer, Kaplan, Laplante, and Salvail addressed
this issue using distributional adversary bounds and an
average-case composition argument~\cite{TQC:BBHKLS17}.
For every fixed
$\eps>0$, they obtained classical protocols with $O(\securityParam)$
expected honest queries for which every quantum eavesdropper making
$O(\securityParam^{1.5-\eps})$ queries has vanishing key recovery probability.
When honest parties can make quantum queries, their separation approaches the quadratic
gap.
Their analysis uses the $\sumArity$-SUM adversary construction
of Belovs and \v Spalek~\cite{ITCS:BelSpa13}.

These results leave a quantitative question for classical honest
parties.
A quantum eavesdropper can always run the classical attack of
Barak and Mahmoody, so the quadratic classical upper bound still
applies.
However, Grover's attack on the original Merkle construction
does not provide a faster attack on every classical protocol.
Can classical parties obtain a nearly quadratic query advantage
against a quantum eavesdropper, or does quantum access always
allow a polynomial improvement over the general classical attack?

\subsection{Our results}

We answer this question tightly, up to logarithmic factors, in the quantum random oracle model.
Alice and Bob
make classical queries to a common random function and exchange classical
messages.
Eve sees the public transcript and can query the same function
in quantum superposition.
We construct protocols in which the honest parties make
$\widetilde O(\securityParam)$ queries, while an adversary must make
at least
$\widetilde\Omega(\securityParam^2)$ quantum queries in order to recover the key\footnote{Security refers to recovery of Bob's output.} with constant probability from the transcript.
Here and in the rest of the paper,
$\widetilde O$ and $\widetilde\Omega$ suppress polylogarithmic factors.
More precisely, our main result can be stated as follows.
\begin{theorem}
There exists a two-message classical protocol using
$\widetilde O(\securityParam)$ oracle queries that establishes
a shared key with probability $1-o(1)$, while any quantum
adversary recovering the key with constant probability
must make $\widetilde\Omega(\securityParam^2)$ oracle queries.
\end{theorem}

This construction has two messages and a small agreement error.
We also give a protocol with
perfect key agreement, following ideas from~\cite{ITC:Mazor25}, which can be
stated as follows.
\begin{theorem}
There exists a three-message classical protocol using
$\widetilde O(\securityParam)$ oracle queries that establishes
a shared key with probability $1$, while any quantum
adversary recovering the key with constant probability
must make $\widetilde\Omega(\securityParam^2)$ oracle queries.
\end{theorem}

Our perfect agreement construction uses one more confirmation message, where on a rare
failure event, Alice and Bob both output the fixed public value $\fallbackKey$.

We also describe a choice of parameters and an implementation with
near-linear honest computation and communication, at the cost of a smaller
quantum-query lower bound.

Fix integers $\innerArity,\outerArity\ge2$.  Under explicit conditions,
we construct a two-message protocol
$\impProtocol_{\innerArity,\outerArity}$ and a three-message protocol
$\perfProtocol_{\innerArity,\outerArity}$.
In both, the honest parties make at most
\(
 O\bigl((\innerArity+\outerArity)\outerArity\,
        \securityParam\log\securityParam\bigr)
\)
classical queries to the oracle.
The first has vanishing agreement error, while the second has perfect agreement.
For every constant $\successProb>0$, recovering Bob's output with
probability at least $\successProb$ requires at least \(\queryBudget\) quantum queries, where
\begin{equation}\label{eq:intro-recovery-scale}
 \queryBudget=\Omega_{\successProb}
       \bigl(\recoveryScale_{\innerArity,\outerArity}(\securityParam)\bigr),
 \qquad
 \recoveryScale_{\innerArity,\outerArity}(\securityParam)
 =\frac{\securityParam^{\innerArity/(\innerArity+1)
                         +\outerArity/(\outerArity+1)}}
        {\sqrt{\innerArity\outerArity}}.
\end{equation}
The precise statements are
\Cref{thm:imperfect-key-recovery,thm:perfect-key-recovery}.

The success probability is averaged over the actual protocol executions,
including the random oracle, honest randomness, and Eve's measurements.
The query budget $\queryBudget$ is a worst-case bound over transcripts and
measurement outcomes.
In particular, any family of Eve algorithms using
$o(\recoveryScale_{\innerArity,\outerArity}(\securityParam))$ queries has
recovery probability tending to zero.

\paragraph{Near-quadratic separation}
The bounds keep their dependence on both arities explicit.
Taking \(\innerArity=\outerArity=\sumArity\) as a function of \(\securityParam\)
gives honest query complexity
$\widetilde O(\securityParam )$
and quantum key-recovery complexity
$\widetilde \Omega(\securityParam^2 )$.
The imperfect protocol's agreement error and the perfect protocol's
probability of outputting $\fallbackKey$ are both
$O(\securityParam^{-3})$.
See~\Cref{thm:uniform} for the precise statement.
For every fixed $\eps>0$, a sufficiently large constant
$\innerArity=\outerArity$ instead gives
$\Omega_{\eps,\successProb}(\securityParam^{2-\eps})$ security with
$O_\eps(\securityParam\log\securityParam)$ honest queries and polynomial
honest running time.
Note that the degree of that polynomial depends on $\eps$.

\paragraph{When honest computation matters}
The near-quadratic query result is near optimal in the query complexity model, but requires expensive local searches.
For a more realistic implementation, take $\innerArity=\outerArity=2$ and slightly modify the protocol, as explained in~\Cref{sec:public-targets}.
Then both the imperfect and perfect
versions have \(O(\securityParam(\log\securityParam))\) queries and \(O(\securityParam(\log^2\securityParam))\) word operations, while any quantum attacker must make \(\Omega_{\successProb}(\securityParam^{4/3})\) queries to recover the key.
This improves the \(\securityParam^{7/6}\) security exponent of the earlier classical construction with near-linear honest computations~\cite{TQC:BBHKLS17,JC:BHKKLS19}.

\subsection{Related work}

\paragraph{Impossibility results in the quantum random-oracle model}
A related line of work allows the honest parties themselves to make quantum
queries, while restricting their communication.
Austrin, Chung, Chung,
Fu, Lin, and Mahmoody proved a quadratic classical-query attack when one
party is classical and agreement is perfect~\cite{C:ACCFLM22}.
For two quantum parties exchanging classical messages, they obtained
polynomial-query classical attacks under their polynomial compatibility
conjecture (PCC).
Bouaziz{-}{-}Ermann, Grilo, Vergnaud, and Vu extended this approach
to a final quantum message under the same conjecture and a restriction on
the receiver's subsequent oracle access~\cite{CiC:BGVV24}.
More recently, Li, Li, Li, and Liu gave an unconditional
$O(\securityParam^5)$ classical-query attack on perfectly agreeing protocols
with arbitrarily many classical messages and a Boolean-valued random
oracle~\cite{LiEtAlManyRound26}.

These attacks use classical oracle queries, even when they attack quantum
honest parties.
Li, Li, Li, and Liu
developed quantum-query attacks using quantum Markov chains, first for
restricted public-key encryption and then for general perfectly correct
quantum public-key encryption with classical
keys~\cite{C:LLLL24,ITCS:LLLL25,C:LLLL26}.
Kitagawa, Nishimaki, Villanyi, and Yamakawa also obtained attacks allowing
imperfect agreement under restrictions on the honest parties' oracle
access and communication~\cite{KitagawaEtAl26}.
Thus several central impossibility arguments use classical queries, while
other arguments do use quantum queries.
These works study whether
\emph{polynomial-query} attacks exist, and do not establish a general
quantum speedup over the quadratic classical attack on classical
protocols.
Our result concerns this quantitative question: no fixed
polynomial improvement over the quadratic exponent is possible for all
classical protocols in the query model.

\paragraph{Perfect agreement}
Perfect agreement requires the parties to output the same key in every
execution, including those involving unlikely oracle collisions.
For classical protocols with this property,
Brakerski, Katz, Segev, and Yerukhimovich~\cite{TCC:BKSY11}
give a classical eavesdropper that always recovers the key using
$O(\securityParam^2)$ oracle queries when each honest party makes at most
$\securityParam$ queries.
Their analysis uses a simple consequence of perfect agreement:
if a candidate execution consistent with the transcript gives the wrong
key, scanning its queries reveals at least one previously unknown query
made by the other party.
This gives a shorter proof than the general quadratic attack of
Barak and Mahmoody~\cite{C:BarMah09}, which also applies to protocols
with agreement errors.

Mazor~\cite{ITC:Mazor25} showed that the quadratic bound remains tight
under perfect agreement, by constructing a two-message protocol with
a quadratic query separation against classical eavesdroppers.
The parties' query sets are chosen to always have exactly one common
input.
Bob checks for additional collisions among their oracle answers.
If the check fails, he informs Alice and both parties output the same
fixed public key.
These executions occur with small probability and are included in the
security analysis.
Thus perfect agreement does not require the key to remain hidden in
every execution.

Mazor's protocol uses two sequential messages: Bob's response depends
on Alice's message.
In comparison, the usual imperfect-agreement Merkle protocol admits
a non-interactive implementation in which both parties send their messages
simultaneously.
Here, ``non-interactive'' refers to this simultaneous-message property,
not to the absence of communication.

\paragraph{From a failed attack to a protocol}

Our starting point was an attempt to improve a classical attack using
quantum search, as we believed that the bound of \(\securityParam^{1.5}\) was optimal.
In attack of perfect protocols from~\cite{TCC:BKSY11}, Eve constructs a possible view of Alice
that agrees with the public transcript and with the oracle answers she
has already learned.
If the key in this view is wrong, the argument
forces a conflict with Bob's actual view on a query that Eve has not yet
learned.
Querying all positions in the candidate view therefore makes
progress, and this operation costs \(\securityParam\) queries.
Repeating this step \(O(\securityParam)\) times leads to a quadratic query cost.

We attempted to replace this scan by a Grover search for an oracle position where
the answer in the simulated view differs from the real oracle answer.
The issue
is that a candidate view can contain many such discrepancies at positions
Bob never queried.
When the candidate key in Eve's simulation is wrong, the classical Eve is guaranteed to query at least one position that Bob queried, even though Eve cannot identify which one it is.
A Grover search may instead return a discrepancy at a position that Bob never queried, which does not guarantee the progress used in the classical analysis.
Thus the progress argument for the classical attack does
not by itself give a quantum speedup.

This obstacle suggested a construction in which useful information is
carried by a relation among several oracle values.
Alice can create such a
relation cheaply, because she can choose the positions.
Bob can recover it
by reading a whole block.
Eve must also determine \emph{which} blocks
matter for the key.
We use a planted sum inside each block to hide a bit,
and another planted sum among the decoded labels to hide the key.
The failed attack is what motivated the design of the protocol.

\paragraph{Discussion and open questions}

The construction separates two tasks that an eavesdropper must solve:
learning information hidden inside blocks and identifying the combination
of blocks that determines the key.
Exact alignment of the planted-sum
operators makes these two costs compose on the protocol's actual
distribution.
The explicit dependence on the arities then brings the
query separation within polylogarithmic factors of the classical
quadratic upper bound.

Several questions remain.
First, can the polylogarithmic loss be removed?
Our growing-arity choice is limited by
$\sumArity^2\securityParam^{-1/(\sumArity+1)}\le1$, and optimizing this choice
alone does not give a constant-factor quadratic lower bound.
Second, can one build a protocol whose honest running time is also linear?
Our protocol is computationally inefficient, as the honest parties must solve subset sum instances.
This issue also existed in previous constructions~\cite{JC:BHKKLS19,TQC:BBHKLS17}.
\Cref{sec:public-targets} gives a near linear time implementation with a smaller security exponent.

Our construction for the imperfect case requires two messages (one from Alice
and one from Bob) which is optimal\footnote{With only one message, Eve can simply simulate the party that receives the message to obtain the key.}, but we ask whether the protocol can be made
non-interactive.
For the perfect case, our protocol requires three messages, and we ask if the same security can be achieved with two messages.

Finally, our security statement concerns recovery of the entire subset
key by a passive eavesdropper.
It does not by itself establish an
indistinguishability-based key-agreement definition.
Obtaining such a
statement requires a separate argument, for example with a
privacy amplification step~\cite{TCC:RenKon05}.

\subsection{AI disclosure}

The authors first considered a general \(O(N^{3/2})\)-query quantum attack on classical key agreement protocol,  %
where the honest parties make \(O(N)\) oracle queries.
The candidate attack was explained to ChatGPT, who identified a counterexample.
We then explored explored different attack approaches, mainly based on Grover search and quantum walks.
For each attack we described, ChatGPT identified a counterexample.

We then asked whether the obstacles encountered in these attempts could instead suggest a protocol achieving a quadratic query separation against quantum eavesdroppers.
In response, ChatGPT proposed the construction developed and analyzed in this paper, which achieves a near-quadratic separation.
ChatGPT also assisted with drafting.
The author takes responsibility for the mathematical claims and proofs.

\paragraph{Acknowledgements}
We thank Alexandru Cojocaru and Quoc-Huy Vu for the helpful discussions.

\section{Technical overview}
\label{sec:overview}

In this technical overview, we will focus on the two-message imperfect-agreement protocol.
Let \(\securityParam \in \mathbb N\) be the security parameter, and fix public parameters \(r,l \in \mathbb N\), \( \outerBits=\lceil(\outerArity+3)\log_2\securityParam\rceil, \innerBits=\lceil(\innerArity+4)\log_2\securityParam\rceil\).
We consider the random oracle model, where all the parties have oracle access to a uniformly random function
\[
 \randomOracle :[\securityParam ]\times[\outerBits ]\times[\securityParam ]
   \longrightarrow\F_2^\innerBits .
\]

Alice samples \(a_1,\dots,a_{\securityParam}\in\F_2^\outerBits\) uniformly at random, which we
call outer labels.
She encodes each bit \(a_i(t)\) using a separate oracle block, consisting of the \(\securityParam\) values \((\randomOracle(i,t,1),\dots,\randomOracle(i,t,\securityParam))\).
Thus each outer label uses \(\outerBits\) blocks, and there are \(\securityParam\outerBits\) blocks in total.
For simplicity, we omit dependence in \((i,t)\) and write \(\randomOracle(j)=\randomOracle(i,t,j)\).

\subsection{Alice's encoding}

For one block, Alice publishes masks $\publicMask_1,\ldots,
\publicMask_{\securityParam}$ and Bob reads the masked values
\(
 Y_j=\randomOracle(j)+\publicMask_j.
\)
To encode zero, Alice chooses every mask uniformly at random and makes no oracle
query.
The masked block is then uniformly random.
To encode one, she chooses a uniformly random subset
$S\in\binom{[\securityParam]}{\innerArity}$, queries its
$\innerArity$ positions, and chooses the masks so that
\[
 \sum_{j\in S}Y_j=0.
\]
For example, after choosing a pivot $p\in S$ and every other mask
uniformly, she sets
\[
 \publicMask_p=\randomOracle(p)
       +\sum_{j\in S\setminus\{p\}}
               \bigl(\randomOracle(j)+\publicMask_j\bigr).
\]
The resulting block is uniform subject to this one equation.
We call
these two distributions the null distribution $\nullDist$ and the planted
distribution $\plantedDist$.
In both cases the public mask is uniform
and independent of the masked block.
See~\Cref{lem:mask-distribution} for a precise statement.
Bob decodes by reading all $\securityParam$ entries and testing whether
any $\innerArity$ distinct positions sum to zero.
The test always returns
one on a planted block.
On a null block, a union bound gives
\[
 \Prb[\text{the test returns one}]
 \le\binom{\securityParam}{\innerArity}2^{-\innerBits}
 \le\securityParam^{-4}.
\]
Thus Alice uses at most $\innerArity$ queries to encode a bit, while Bob
uses $\securityParam$ queries to decode it.
Note that Bob's subsequent search uses
only the values already read and requires no further oracle queries, but can be computationally inefficient.

\subsection{Bob's message and the shared key}
After encoding all \(\securityParam\outerBits\) bits, Alice sends the mask table to Bob.
Bob chooses a uniformly random subset
$\sharedKey\in\binom{[\securityParam]}{\outerArity}$, reads the
$\outerArity\outerBits$ blocks needed to decode its labels, and obtains
$\widehat a_i$ for $i\in\sharedKey$.
He sends Alice the message
\[
 z=\sum_{i\in\sharedKey}\widehat a_i.
\]
Bob outputs $\sharedKey$ as his key.
Alice already knows every original label, so
she searches locally for the unique subset $J$ of size $\outerArity$ with
\[
 \sum_{j\in J}a_j=z.
\]
She outputs this subset when it exists and is unique. Otherwise, she aborts.

If Bob decoded correctly, then $\sharedKey$ is a solution.
Without
conditioning on this decoding event, the probability that another subset
has the same sum as $\sharedKey$ in the original labels is at most
$\binom{\securityParam}{\outerArity}2^{-\outerBits}
\le\securityParam^{-3}$.
Outside these two bad events, Alice's unique
solution is $\sharedKey$.
Adding their probabilities bounds the total
probability of failure to agree by
\[
 \outerArity\outerBits\securityParam^{-4}+\securityParam^{-3}.
\]
Alice makes at most $\innerArity\securityParam\outerBits$ queries, and
Bob makes $\outerArity\outerBits\securityParam$.
The two costs come
from different tasks: Alice prepares many cheap encodings, whereas Bob
fully reads a few of them.

\subsection{Security argument}

\paragraph{Two levels of hidden sums}
Consider an ideal experiment in which Bob sends
$z=\sum_{i\in\sharedKey}a_i$ using the original labels.
The real and ideal
transcripts differ only if one of his decoded bits is wrong.
The
probability of this event is at most
$\outerArity\outerBits\securityParam^{-4}$.
For a fixed public $z$, the ideal label vector has exactly the planted
$\outerArity$-SUM distribution: choose a hidden subset and sample labels
conditioned on their sum being $z$.

If Eve could read a label directly, this outer problem would already be
hard.
Here its bits are themselves hidden in planted-sum encodings.
The two relevant lower-bound scales are
\[
 \frac{\securityParam^{\outerArity/(\outerArity+1)}}
      {\sqrt{\outerArity}}
 \quad\text{and}\quad
 \frac{\securityParam^{\innerArity/(\innerArity+1)}}
      {\sqrt{\innerArity}}.
\]
Their product is the scale in~\eqref{eq:intro-recovery-scale}.
However, this multiplication needs a proof: Eve can query different blocks in
superposition and does not necessarily decode labels one at a time.

\paragraph{Preserving the protocol's distributions}
We use the average-case adversary method, which bounds the number of
quantum queries needed to distinguish a planted distribution from a null
distribution.
The method associates an operator with these distributions
and compares its size with the change that one query can cause.
Our main technical step strengthens the norm analysis of the known
planted $\sumArity$-SUM operator.
Uniformly in the target sum, we prove
\[
 \|\advMatrix_{\sumArity,\securityParam}\|
     =\securityParam^{\sumArity/(\sumArity+1)},
 \qquad
 \|\advMatrix_{\sumArity,\securityParam}\circ\queryDiff_j\|
     \le9\sqrt{\sumArity},
\]
whenever $\sumArity^2\securityParam^{-1/(\sumArity+1)}\le1$.
Here the operation $\circ\queryDiff_j$ keeps the matrix entries whose oracle answers
at position $j$ differ.
Moreover, the normalized states whose amplitudes are the square roots of
the planted and null probabilities satisfy
\[
 \advMatrix\distKet{\nullDist}
     =\|\advMatrix\|\distKet{\plantedDist},
 \qquad
 \advMatrix^{\adj}\distKet{\plantedDist}
     =\|\advMatrix\|\distKet{\nullDist}.
\]
These exact identities let us apply standard adversary
composition~\cite{STOC:HoyLeeSpa07,BelovsLee20} while preserving the
distributional quantity needed in the average-case bound.
An approximate identity could lose a factor in each of the
$\securityParam\outerBits$ encoded bits.
Exact alignment avoids that
accumulated loss.
The composition norm theorem itself is a result from
the literature.
In this work, we verify the alignment and bounds needed for
these planted distributions.%

\paragraph{Perfect agreement upgrade}
To construct a perfect protocol, we encode each bit using a pair of blocks:
\[
 \text{zero}:\quad\plantedDist\otimes\nullDist,
 \qquad
 \text{one}:\quad\nullDist\otimes\plantedDist.
\]
At least one block contains a relation.
If exactly one does, Bob learns
the bit with certainty.
If both do, he requests the fallback key
$\fallbackKey$.
Alice's final confirmation message, which is the third message of the protocol, confirms a unique outer solution or
announces the same fallback key.
Consequently the parties always agree.
The fallback probability is at most
$\outerArity\outerBits\securityParam^{-4}+\securityParam^{-3}$, and Eve knows the key on these executions.
The paired construction changes the query costs only by constant factors
and preserves the inner adversary ratio.

\paragraph{Organization of the paper}

In~\Cref{sec:normalization} we prove the exact norm and explicit query-filter bound,
and in~\Cref{sec:composition} we apply these results to planted-sum
distributions and to paired encodings.
In~\Cref{sec:imperfect-protocol} we describe the two-message protocol and prove
its correctness and security.
In~\Cref{sec:perfect} we describe the perfect three-message protocol, along with its security.
Finally, in~\Cref{sec:parameters} we derive the near-quadratic
parameter choice and an efficient implementation.

%% file: tex/preliminaries2.tex
\section{Preliminaries}
\label{sec:preliminaries}

We use standard notation from quantum information.
For background on
quantum query algorithms and the adversary method, we refer to~\cite[Sections~1 and~5]{Hamoudi25}.
The distributional bound and planted-sum operator used in this section come
from~\cite{TQC:BBHKLS17}, building on~\cite{ITCS:BelSpa13}.
The matrix composition identities
come from the adversary composition literature~\cite{STOC:HoyLeeSpa07,BelovsLee20}.
Our exact normalization and uniform bounds are proved in~\Cref{sec:normalization}.
We apply the composition identities to our
chosen distributions in~\Cref{sec:composition}.

\subsection{Notation and the query model}

We write $[\sumSize ]=\{1,\ldots,\sumSize \}$ and $\binom{[\sumSize ]}\sumArity $ for the collection of
$\sumArity $-element subsets of $[\sumSize ]$.
For a finite label set $\cX$, let
\[
 \labelSpace{\cX}=\operatorname{span}\{\ket{x}:x\in\cX\},
 \qquad \braket{x}{y}=\mathbf1[x=y].
\]
All Hilbert spaces are finite-dimensional and complex.  A matrix with rows
indexed by $\cX$ and columns indexed by $\cY$ is the linear operator
\[
 A:\labelSpace{\cY}\longrightarrow\labelSpace{\cX},
 \qquad
 A=\sum_{x\in\cX,\,y\in\cY}\matrixel{x}{A}{y}\ketbra{x}{y}.
\]
It is not necessarily a square matrix, Hermitian, or unitary.
We write $A^{\adj}$ for its
adjoint and $\identity_{\cX}$ for the identity on $\labelSpace{\cX}$.
We suppress the subscript when the space is clear.
The operator norm is
\[
 \|A\|=\max_{\|\ket{v}\|=1}\|A\ket{v}\|
       =\max_{\|\ket{u}\|=\|\ket{v}\|=1}
          |\matrixel{u}{A}{v}|.
\]
Here $\|\ket{v}\|=\sqrt{\braket{v}{v}}$ is the vector norm.
The notation $A\circ B$ denotes the entrywise product in the stated label
bases:
\[
 \matrixel{x}{A\circ B}{y}
 =\matrixel{x}{A}{y}\matrixel{x}{B}{y}.
\]
It differs from operator multiplication $AB$ and from the tensor product
$A\otimes B$.
We use $\|A\otimes B\|=\|A\|\|B\|$ and the triangle
inequality.
An isometry $V$ satisfies $V^{\adj} V=\identity$.
For a Hermitian operator $K$ and any orthonormal basis $\{\ket{u}\}$,
\begin{equation}\label{eq:column-sum-norm}
 \|K\|\le\max_v\sum_u|\matrixel{u}{K}{v}|.
\end{equation}
Indeed, the spectral norm is at most the square root of the product of
the maximum absolute row sum and the maximum absolute column sum.
These two sums agree for a Hermitian matrix.

Unless a base is displayed, logarithms are natural.
The notations
$\widetilde O$ and $\widetilde\Omega$ suppress factors polynomial in
$\log \securityParam $.
A probability is \emph{vanishing} if it tends to zero.  It is
\emph{negligible in $\securityParam $} if it is smaller than $\securityParam ^{-c}$ for every fixed
$c>0$, for all sufficiently large $\securityParam $.
Constants in $O_\sumArity (1)$ may depend on $\sumArity $, whereas a stated
universal constant does not.

An oracle input is a string $x=(x_1,\ldots,x_\sumSize )$ over a finite abelian group
$\sumGroup $.
A quantum query acts as follows
\[
 O_x\ket{j,b,w}=\ket{j,b+x_j,w}.
\]
The query index $j$ may be in superposition.  Local computation, memory,
and classical communication are free.  A $\queryBudget $-query algorithm uses at most
$\queryBudget $ queries on every branch.
We allow a dummy index whose oracle value is
always zero, so a shorter branch can be padded.
We represent classical randomness and intermediate measurements using
extra registers.
We may then perform all measurements at the end,
without changing the query count.

Our distributions are sometimes defined on \emph{labelled inputs}.  A label $u$
specifies an oracle string $x(u)$, but different labels may specify the same
string.
An input label is an index used in the analysis and may contain auxiliary information.
The algorithm receives access only to $x(u)$, not to the label.
For row and column labels $u,v$, write
\[
 \matrixel{u}{\queryDiff_j}{v}=\mathbf 1[x_j(u)\ne x_j(v)].
\]
Thus $\advMatrix \circ\queryDiff _j$ retains the comparisons on which a query at
position $j$ can detect a difference.
We use the superscripts
$\mathrm{word}$ and $\mathrm{bit}$ when both types of query position occur.

For a probability distribution $\plantedDist$ on labels, define its
\emph{square-root state} by
\[
 \distKet{\plantedDist}
 =\sum_{u\in\supp(\plantedDist)}\sqrt{\plantedDist[u]}\ket{u},
 \qquad
 \braket{\sqrtDist_{\plantedDist}}{\sqrtDist_{\plantedDist}}
 =\sum_u\plantedDist[u]=1.
\]
Thus $\braket{u}{\sqrtDist_{\plantedDist}}=\sqrt{\plantedDist[u]}$.
This is a pure state that is used only in the analysis.
The algorithm is not given
coherent access to this state or to its label register.
Independent distributions give tensor products:
\[
 \distKet{\plantedDist\otimes\nullDist}
 =\distKet{\plantedDist}\otimes\distKet{\nullDist}.
\]

In the rest of the paper, we write \(\securityParam\) for the security parameter.
In~\Cref{sec:normalization}, we write \(\sumSize\) for the length of a single sum instance and \(\sumArity\) for the size of its planted sum.
For the protocols in~\Cref{sec:imperfect-protocol}, we write \(\outerSize,\outerArity\) for the lengths and subset size for the \emph{outer} instances, that is Bob's planted instance, and \(\innerSize,\innerArity\) for the lengths and subset size of the inner blocks, that is Alice's planted instances.
We write \(\outerBits\) for the number of bits in an outer word and \(\innerBits\) for the number of bits in an inner oracle value.
There are \(\numBlocks=\outerSize \outerBits\) outer bits, indexed by \(p = (i,t)\).
In the protocols of~\Cref{sec:imperfect-protocol}, we take \(\outerSize=\innerSize=\securityParam\).

\subsection{The distributional adversary theorem}

To prove a query lower bound, we choose an adversary operator.
Its matrix
elements assign weights to pairs of inputs.  These weights define a
progress measure for distinguishing the two distributions.  The weights
may be negative or complex.
Following the
register formulation of~\cite[Section~5.1]{Hamoudi25}, we introduce a
label register used only in the analysis.
The algorithm cannot access it.
If $\ket{\psi_u^t}$ is its state after $t$
queries on label $u$, define, for $R\in\{\plantedDist,\nullDist\}$,
\[
 \ket{\Psi_R^t}
 =\sum_{u\in\supp(R)}\sqrt{R[u]}\ket{u}\otimes\ket{\psi_u^t}.
\]
The distributional method tracks the weighted overlap
\[
 W_t=\operatorname{Re}
 \matrixel{\Psi_{\plantedDist}^t}
 {\advMatrix\otimes\identity}{\Psi_{\nullDist}^t}.
\]
Initially the algorithm's state is independent of $u$, so
$W_0=\operatorname{Re}(\distBra{\plantedDist}\advMatrix\distKet{\nullDist})$.
Unitaries that do not depend on the input leave this overlap unchanged.
The theorem below bounds how much each query can change it.  It also
relates the final overlap to the two output distributions.

For acceptance probabilities $\acceptProb _\plantedDist ,\acceptProb _\nullDist \in[0,1]$, define
\begin{equation}\label{eq:tau-definition}
 \outputOverlap (\acceptProb _\plantedDist ,\acceptProb _\nullDist )
 =\sqrt{\acceptProb _\plantedDist \acceptProb _\nullDist }+\sqrt{(1-\acceptProb _\plantedDist )(1-\acceptProb _\nullDist )}.
\end{equation}
This is the inner product of the two normalized states
$\sqrt{\acceptProb_R}\ket{1}+\sqrt{1-\acceptProb_R}\ket{0}$ for
$R\in\{\plantedDist,\nullDist\}$.
In particular, $0\le\outputOverlap \le1$ and
\[
 1-\outputOverlap
 =\frac12(\sqrt{\acceptProb _\plantedDist }-\sqrt{\acceptProb _\nullDist })^2
  +\frac12(\sqrt{1-\acceptProb _\plantedDist }-\sqrt{1-\acceptProb _\nullDist })^2.
\]
Identical output distributions give $\outputOverlap =1$, while perfect distinction
gives $\outputOverlap =0$.

\begin{theorem}%
\label{thm:average-adv}
Let $\plantedDist ,\nullDist $ be distributions on labelled oracle inputs, and let $\advMatrix $ have
rows indexed by $\supp(\plantedDist )$ and columns indexed by $\supp(\nullDist )$.  Set
\[
 D=\max_j\|\advMatrix \circ\queryDiff _j\|.
\]
If an algorithm makes at most $\queryBudget $ quantum queries and accepts with average
probabilities $\acceptProb _\plantedDist ,\acceptProb _\nullDist $ under $\plantedDist ,\nullDist $, respectively, then
\begin{equation}\label{eq:adversary-explicit}
 2\queryBudget D\ge
 \operatorname{Re}(\distBra{\plantedDist}\advMatrix \distKet{\nullDist} )
 -\outputOverlap (\acceptProb _\plantedDist ,\acceptProb _\nullDist )\|\advMatrix \|.
\end{equation}
\end{theorem}

The precise bound above is the distributional adversary theorem of~\cite[full version, Theorem~4 and Claims~1--3]{TQC:BBHKLS17}, where the explicit factor $2$
comes from the one-query estimate in Claim~3.
The cited theorem allows rectangular matrices by adding zero rows and
columns, as explained immediately after its statement.
Labels are treated as input indices, with oracle answers determined by
their associated strings.
Different labels may therefore give the same oracle.

The probabilities in this theorem are averaged over the input
distributions.
The query budget $\queryBudget $ is a worst-case budget over executions.
These are different uses of ``average case'' and ``worst case''.
For our operators, the initial overlap equals the norm.  The numerator
then becomes
$(1-\outputOverlap )\|\advMatrix \|$.
A constant distinguishing advantage therefore gives a lower bound
proportional to the adversary ratio.

\subsection{Planted sum distributions}

Fix a finite abelian group $\sumGroup $ of size $\groupSize $, a target $w\in \sumGroup $, and
$\sumSize \ge \sumArity \ge2$.  Write
\[
 \plantedSpace _S^w=\{x\in \sumGroup ^\sumSize :\textstyle\sum_{i\in S}x_i=w\}
 \quad(S\in\tbinom{[\sumSize ]}\sumArity ).
\]
The null and planted labelled sample spaces are
\[
 \nullSpace _{\sumArity ,\sumSize }=\sumGroup ^\sumSize ,
 \qquad
 \plantedSpace _{\sumArity ,\sumSize }^w=\{(S,x):S\in\tbinom{[\sumSize ]}\sumArity ,\ x\in\plantedSpace _S^w\}.
\]
Let $\nullDist _{\sumArity ,\sumSize }$ and
$\plantedDist _{\sumArity ,\sumSize }^w$ be the uniform distributions on these
spaces. A planted sample is generated by choosing $S$ uniformly and then
choosing $x$ uniformly subject to the equation on $S$. When only \(x\) is
relevant, we write \(x \leftarrow \plantedSpace _{\sumArity ,\sumSize }^w\) as a
shorthand for sampling
\((S,x) \leftarrow \plantedSpace _{\sumArity ,\sumSize }^w\) and retaining \(x\).
In the adversary analysis, we keep the full pair \((S,x)\) as an input label.
The algorithm has oracle access only to \(x\) and is not given \(S\).

We call $S$ a \emph{certificate}: its $\sumArity$ positions witness the
relation $\sum_{i\in S}x_i=w$.  Reading those positions is enough to verify
that this particular relation holds.
There are $\groupSize ^{\sumSize -1}$ such strings for each
$S$: choose one position in $S$ and assign all other coordinates freely; the
sum equation determines the remaining coordinate.
Thus
\[
 \nullDist _{\sumArity ,\sumSize }[x]=\groupSize ^{-\sumSize },\qquad
 \plantedDist _{\sumArity ,\sumSize }^w[S,x]=\binom \sumSize \sumArity ^{-1}\groupSize ^{-(\sumSize -1)}.
\]
A string with several solutions
occurs under several planted labels.
The resulting distribution on strings
is weighted by the number of solutions; it is not the uniform distribution
on strings having at least one solution.
Also, a null sample is not
conditioned to have no solution.  For later use, a union bound gives
\begin{equation}\label{eq:generic-accidental-sum}
 \Prb_{x\leftarrow \nullDist _{\sumArity ,\sumSize }}
       [\exists S:\textstyle\sum_{i\in S}x_i=w]
 \le\binom \sumSize \sumArity /\groupSize .
\end{equation}
Indeed, for each fixed $S$, its sum is uniform in $\sumGroup $, so its probability of being
$w$ is exactly $1/\groupSize $.

\subsection{Fourier description of the planted \texorpdfstring{$\sumArity$}{\sumArity}-SUM matrix}

We now define the adversary operator used to analyze the protocol.
A character of $\sumGroup $ is a map
$\character :\sumGroup \to\mathbb C$ with $|\character (x)|=1$ and
$\character (x+y)=\character (x)\character (y)$.
The characters form a group $\widehat \sumGroup $
of size $\groupSize $.
We write its operation additively: $\character _{a+b}=\character _a\character _b$, and $0$
denotes the constant character.
Character orthogonality says
\[
 \frac1\groupSize \sum_{x\in \sumGroup }\overline{\character _a(x)}\character _b(x)
 =\mathbf 1[a=b].
\]

On one coordinate, define
\[
 \fourierKet{a}
 =\frac1{\sqrt{\groupSize}}\sum_{x\in\sumGroup}\character_a(x)\ket{x}
 \qquad(a\in\widehat\sumGroup).
\]
For $v\in\widehat\sumGroup^{\sumSize}$, the corresponding Fourier basis state is
\[
 \fourierKet{v}
 =\bigotimes_{i=1}^{\sumSize}\fourierKet{v_i}
 =\groupSize^{-\sumSize/2}\sum_{x\in\sumGroup^{\sumSize}}
       \left(\prod_{i=1}^{\sumSize}\character_{v_i}(x_i)\right)\ket{x}.
\]
Character orthogonality makes these states an orthonormal basis.
Write $\supp(v)=\{i:v_i\ne0\}$ and $\wt(v)=|\supp(v)|$.
For $\sumGroup=\F_2^{\innerBits}$, one may simply use
$\character_a(x)=(-1)^{a\cdot x}$.

On one coordinate, let
\[
 \fourierProjector_0=\ketbra{\fourierVector_0}{\fourierVector_0},
 \qquad
 \fourierProjector_1=\identity-\fourierProjector_0.
\]
The first projector has every matrix element equal to $1/\groupSize$;
the second projects onto the nonconstant Fourier states.
For $U\subseteq[\sumSize]$, let
\[
 \fourierProjector_U
 =\bigotimes_{i=1}^{\sumSize}\fourierProjector_{\mathbf1[i\in U]},
 \qquad
 \fourierProjector_U\fourierKet{v}
 =\mathbf1[\supp(v)=U]\fourierKet{v}.
\]
To express restriction to the planted rows as an operator, define
\[
 \restrictionOp_S^w:\labelSpace{\sumGroup^{\sumSize}}
       \longrightarrow\labelSpace{\plantedSpace_S^w},
 \qquad
 \restrictionOp_S^w=\sum_{x\in\plantedSpace_S^w}\ketbra{x}{x}.
\]
In this formula the ket belongs to the restricted output space and the
bra to the full input space: $\restrictionOp_S^w$ retains precisely the
allowed computational-basis components.
For a nonnegative coefficient sequence $\sumCoeff$, set
\begin{equation}\label{eq:planted-sum-matrix}
 \begin{aligned}
 \sumBlock_S(\sumCoeff)
 &=\sqrt{\groupSize}\,\restrictionOp_S^w
       \sum_{U:\,S\not\subseteq U}\sumCoeff_{|U|}\fourierProjector_U,\\
 \advMatrix(\sumCoeff)
 &=\sum_{S\in\binom{[\sumSize]}{\sumArity}}
       \ket{S}\otimes\sumBlock_S(\sumCoeff).
 \end{aligned}
\end{equation}
The output space is the orthogonal direct sum
$\bigoplus_S\bigl(\ket{S}\otimes\labelSpace{\plantedSpace_S^w}\bigr)$.
Thus a row label is still $(S,x)$, and the last formula is exactly the
vertical stack of the operators $\sumBlock_S$.
This is the construction of
\cite[full version, Section~5, equations~(8)--(10)]{TQC:BBHKLS17}, based
on~\cite{ITCS:BelSpa13}.  Restricting to $\groupSize^{\sumSize-1}$ of the $\groupSize^{\sumSize}$ rows
gives each restricted Fourier state norm $1/\sqrt{\groupSize}$.
The factor $\sqrt{\groupSize}$ makes these states have norm one.

For the rest of the paper, unless another sequence is specified, take
\begin{equation}\label{eq:alpha-definition}
 \sumScale =\sumScale _{\sumArity ,\sumSize }=\sumSize ^{\sumArity /(\sumArity +1)},\qquad
 \sumCoeff _a=\binom \sumSize \sumArity ^{-1/2}(\sumScale -a)_+,
 \qquad(t)_+=\max\{t,0\},
\end{equation}
and write $\advMatrix _{\sumArity ,\sumSize }=\advMatrix (\sumCoeff )$.  The target $w$ and group $\sumGroup $
are omitted from this notation.  The linear decrease in $a$ will control
both the full norm and the norm after one query.

We use the following identities from~\cite[full version, Section~5, equations~(11)--(12) and Claim~6]{TQC:BBHKLS17},
written in character notation.
For a nonzero target, a common shift
can introduce a phase; the absolute value in~\eqref{eq:fourier-block}
accounts for it.

\begin{lemma}%
\label[lemma]{lem:planted-sum-identities}
Write $\restrictedKet{S}{v}=\sqrt{\groupSize}\,\restrictionOp_S^w\fourierKet{v}$.
This is a unit vector, and
\begin{equation}\label{eq:basic-block-action}
 \sumBlock _S(\sumCoeff )\fourierKet{v}=
 \begin{cases}
  \sumCoeff _{\wt(v)}\restrictedKet{S}{v},&S\not\subseteq\supp(v),\\
  0,&S\subseteq\supp(v).
 \end{cases}
\end{equation}
Two vectors $\restrictedKet{S}{u},\restrictedKet{S}{v}$ are orthogonal unless $v$ is obtained
from $u$ by adding one common character index on every coordinate of $S$
and leaving the other coordinates unchanged.
In that case their inner
product has modulus one.  In particular,
\begin{equation}\label{eq:fourier-block}
 |\fourierBra{u}\sumBlock_S^{\adj}\sumBlock_S\fourierKet{v}|=
 \begin{cases}
 \sumCoeff _{\wt(u)}\sumCoeff _{\wt(v)},&
 \begin{array}{l}S\not\subseteq\supp(u),\ S\not\subseteq\supp(v),\\
             \text{$u,v$ differ by a common shift on $S$},\end{array}\\[1mm]
 0,&\text{otherwise}.
 \end{cases}
\end{equation}
For the coefficients in~\eqref{eq:alpha-definition},
\begin{equation}\label{eq:published-overlap}
 \advMatrix _{\sumArity ,\sumSize }\distKet{\nullDist _{\sumArity ,\sumSize }}=\sumScale \distKet{\plantedDist _{\sumArity ,\sumSize }^w},
 \qquad
 \distBra{\plantedDist _{\sumArity ,\sumSize }^w}\advMatrix _{\sumArity ,\sumSize }\distKet{\nullDist _{\sumArity ,\sumSize }}=\sumScale .
\end{equation}
\end{lemma}

For fixed $\sumArity $,~\cite{TQC:BBHKLS17} also prove
\begin{equation}\label{eq:published-filter}
 \|\advMatrix _{\sumArity ,\sumSize }\|
 \le(1+O_\sumArity (\sumSize ^{-1/(\sumArity +1)}))\sumScale ,
 \qquad
 \|\advMatrix _{\sumArity ,\sumSize }\circ\queryDiff _i^{\mathrm{word}}\|=O_\sumArity (1).
\end{equation}
These estimates are our starting point.  Our proofs do not use the
constants hidden in the asymptotic notation.
We prove the
exact norm $\sumScale $, the adjoint alignment, and the explicit bound $9\sqrt{\sumArity }$ in~\Cref{sec:normalization}.

\subsection{Matrix composition}

Composition of adversary bounds is a standard result, see~\cite[Proposition~49]{Hamoudi25} for its statement for Boolean functions.
We use the matrix form below.
The norm identity is proved in~\cite[Lemma~16]{STOC:HoyLeeSpa07} and~\cite[Lemma~21]{BelovsLee20}.
The query-filter bound is proved in~\cite[proof of Theorem~22,
equation~(8) and Claim~23]{BelovsLee20}.
We state both conclusions for
rectangular operators and labelled oracle inputs.

\begin{lemma}
  \label[lemma]{lem:completed-upper}
Let $\numBlocks\ge1$.  For each $p\in[\numBlocks]$, let
$D_p:\mathcal H_{p,0}\to\mathcal H_{p,1}$ and $\rho_p=\|D_p\|$.
Complete it by setting
\[
 D_p^{1,0}=D_p,\quad D_p^{0,1}=D_p^{\adj},\quad
 D_p^{b,b}=\rho_p\identity_{\mathcal H_{p,b}}
 \quad(b\in\{0,1\}).
\]
Let $B:\labelSpace{\cY}\to\labelSpace{\cX}$, where each label $x$ or $y$
carries a bit string $(x_1,\ldots,x_{\numBlocks})$ or
$(y_1,\ldots,y_{\numBlocks})$.  Different labels may carry the same string.
Define
\[
 K=\sum_{x\in\cX,\,y\in\cY}\matrixel{x}{B}{y}\ketbra{x}{y}
       \otimes\bigotimes_{p=1}^{\numBlocks}D_p^{x_p,y_p}.
\]
This maps
$\bigoplus_y(\ket{y}\otimes\bigotimes_p\mathcal H_{p,y_p})$
to $\bigoplus_x(\ket{x}\otimes\bigotimes_p\mathcal H_{p,x_p})$.
Equivalently, embed each $\mathcal H_{p,b}$ in
$\mathcal H_{p,0}\oplus\mathcal H_{p,1}$ and extend every operator by zero.
These embeddings do not change the operator norm.  Then
\begin{equation}\label{eq:composition-upper}
 \|K\|=\|B\|\prod_{p=1}^{\numBlocks}\rho_p.
\end{equation}

Suppose also that each basis label $u$ in $\mathcal H_{p,b}$ specifies
an oracle string, whose answer at position $j$ is denoted by $u(j)$.
Define the outer bit filter and the inner query filter by
\[
 \matrixel{x}{\queryDiff_p^{\rm bit}}{y}=\mathbf1[x_p\ne y_p],
 \qquad
 \matrixel{u}{\queryDiff_{p,j}^{\rm in}}{v}
   =\mathbf1[u(j)\ne v(j)],
\]
where the second filter has the row and column label sets of $D_p$.
For tuples $u=(u_1,\ldots,u_{\numBlocks})$ and
$v=(v_1,\ldots,v_{\numBlocks})$, the physical query filter is
\[
 \matrixel{x,u}{\queryDiff_{p,j}}{y,v}
   =\mathbf1[u_p(j)\ne v_p(j)].
\]
Then
\begin{equation}\label{eq:matrix-composition-filter}
 \|K\circ\queryDiff_{p,j}\|
 \le\|B\circ\queryDiff_p^{\rm bit}\|
       \|D_p\circ\queryDiff_{p,j}^{\rm in}\|
       \prod_{\substack{r\in[\numBlocks]\\r\ne p}}\rho_r.
\end{equation}
An empty product equals $1$.
Different labels may specify the same
oracle string.
\end{lemma}

\begin{proof}
We use the norm identity for completed
matrix composition from~\cite[Lemma~16]{STOC:HoyLeeSpa07}.
Specifically, if $\Gamma$ is a Hermitian adversary matrix indexed by
distinct bit strings, and the inner Hermitian adversary matrices have
off-diagonal blocks $E_1,\ldots,E_m$, then their completed composition
satisfies
\begin{equation}\label{eq:standard-composition-norm}
 \bigl\|\operatorname{Comp}(\Gamma;E_1,\ldots,E_m)\bigr\|
 =\|\Gamma\|\prod_{r=1}^{m}\|E_r\|.
\end{equation}
Here the completed blocks associated with $E_r$ are
\[
 \begin{pmatrix}
  \|E_r\|\identity&E_r\\
  E_r^{\adj}&\|E_r\|\identity
 \end{pmatrix}.
\]
The result above allows complex Hermitian matrices and different
inner matrices in different coordinates.
See also~\cite[Lemma~21]{BelovsLee20} for the real symmetric
matrix formulation.
We apply~\eqref{eq:standard-composition-norm} twice, after verifying
the required identifications.

For a rectangular operator $W$, define its Hermitian dilation by
\[
 \mathcal D(W)=
 \begin{pmatrix}
  0&W\\
  W^{\adj}&0
 \end{pmatrix}.
\]
Since
\[
 \mathcal D(W)^2=WW^{\adj}\oplus W^{\adj}W,
 \qquad
 \|\mathcal D(W)\|=\|W\|,
\]
passing to this dilation preserves the norm.

Choose an injective map
$c:\cX\sqcup\cY\to\{0,1\}^{a}$ and represent each outer input label $z$ by
\[
 \iota(z)=(z_1,\ldots,z_{\numBlocks},c(z)).
\]
Index $\mathcal D(B)$ by these distinct representations, and extend
it by zero to all strings in $\{0,1\}^{\numBlocks+a}$.
Denote the resulting matrix by $\widehat B$.
It is a Hermitian adversary matrix for a Boolean function that is
zero on $\iota(\cX)$ and one on $\iota(\cY)$; the function can be
defined arbitrarily on unused strings.

Compose $\widehat B$ with $D_1,\ldots,D_{\numBlocks}$ in the original
coordinates and with $[1]$ in each auxiliary coordinate.
All four completed blocks associated with $[1]$ equal $[1]$.
Thus the auxiliary coordinates leave the entries unchanged, and,
after reordering the basis, the composed operator is
\[
 \widehat K=\mathcal D(K)\oplus0.
\]
This identification follows directly from the definition of the
completed composition.

We now apply the cited norm identity
\eqref{eq:standard-composition-norm} to $\widehat B$ and these
$\numBlocks+a$ inner matrices. It gives
\[
 \|K\|
 =\|\widehat K\|
 =\|\widehat B\|\prod_{r=1}^{\numBlocks}\rho_r
 =\|B\|\prod_{r=1}^{\numBlocks}\rho_r,
\]
where $\rho_r=\|D_r\|$ and the auxiliary factors have norm one.
This proves the norm identity in the statement.

Fix a physical query $(p,j)$ and write
\[
 B'=B\circ\queryDiff_p^{\rm bit},
 \qquad
 F_r=
 \begin{cases}
  D_p\circ\queryDiff_{p,j}^{\rm in},&r=p,\\
  D_r,&r\ne p.
 \end{cases}
\]
Complete the matrices $F_r$ using $\|F_r\|\identity$ for their
equal-bit blocks.
We claim that
\begin{equation}\label{eq:filtered-composition-identity}
 K\circ\queryDiff_{p,j}
 =
 \sum_{x\in\cX,\,y\in\cY}
 \matrixel{x}{B'}{y}\ketbra{x}{y}
 \otimes
 \bigotimes_{r=1}^{\numBlocks}F_r^{x_r,y_r}.
\end{equation}

This is an entrywise identity, which we verify directly.
If $x_p=y_p$, the original local factor is $\rho_p\identity$.
Its nonzero entries connect identical labels, whose oracle answers
agree, so the physical query filter removes them.
The right-hand side also vanishes because
$\matrixel{x}{B'}{y}=0$.
In particular, the changed equal-bit completion of $F_p$ contributes
nothing.
If $x_p\ne y_p$, the outer filter leaves the entry of $B$ unchanged,
and filtering the local factor gives $F_p$ or $F_p^{\adj}$.
All other factors remain unchanged.
This proves~\eqref{eq:filtered-composition-identity}.

Finally, apply the rectangular norm identity proved above to
$B'$ and $F_1,\ldots,F_{\numBlocks}$.
This gives
\[
 \|K\circ\queryDiff_{p,j}\|
 \le
 \|B\circ\queryDiff_p^{\rm bit}\|\,
 \|D_p\circ\queryDiff_{p,j}^{\rm in}\|
 \prod_{\substack{r=1\\r\ne p}}^{\numBlocks}\rho_r.
\]
The entrywise verification only requires identical labels to have
identical oracle answers. Distinct labels may therefore specify
the same oracle string.
\end{proof}

%% file: tex/proof2.tex
\section{Exact normalization of the average-case \texorpdfstring{$\sumArity$}{\sumArity}-SUM matrix}
\label{sec:normalization}

In this section, we prove the three properties needed for composition: the full
norm is exactly $\sumScale =\sumSize ^{\sumArity /(\sumArity +1)}$, the two distribution vectors are singular vectors with singular value $\sumScale$, and every query-filtered norm is at most $9\sqrt{\sumArity }$.
The first property prevents a loss at each encoded bit.
The last makes
the bound useful even when $\sumArity $ grows with $\sumSize $.

\subsection{The exact norm and distribution-vector alignment}

\begin{theorem}[Exact norm]\label{thm:exact-norm}
Let $\sumSize \ge \sumArity \ge2$ and write $\sumScale =\sumSize ^{\sumArity /(\sumArity +1)}$.  If
\[
  \frac{\sumArity ^2\sumScale }{\sumSize }\le1,
\]
then the planted $\sumArity $-SUM matrix $\advMatrix _{\sumArity ,\sumSize }$ satisfies
\[
  \|\advMatrix _{\sumArity ,\sumSize }\|=\sumScale .
\]
Also,
\begin{align}
 \advMatrix _{\sumArity ,\sumSize }\distKet{\nullDist _{\sumArity ,\sumSize }}
   &=\sumScale \distKet{\plantedDist _{\sumArity ,\sumSize }^{w}},\label{eq:right-align}\\
 \advMatrix _{\sumArity ,\sumSize }^{\adj}\distKet{\plantedDist _{\sumArity ,\sumSize }^{w}}
   &=\sumScale \distKet{\nullDist _{\sumArity ,\sumSize }}.\label{eq:left-align}
\end{align}
\end{theorem}

\begin{proof}
In this proof, write
\[
 \advMatrix=\advMatrix_{\sumArity,\sumSize},\qquad
 \plantedDist=\plantedDist_{\sumArity,\sumSize}^{w},\qquad
 \nullDist=\nullDist_{\sumArity,\sumSize}.
\]
The certificate registers are orthogonal.
Therefore the Gram operator is
\[
 \gramMatrix=\advMatrix^{\adj}\advMatrix
 =\sum_{S\in\binom{[\sumSize]}{\sumArity}}\sumBlock_S^{\adj}\sumBlock_S.
\]

We will bound every absolute column sum of $\gramMatrix $ by $\sumScale ^2$ in the Fourier
basis.
By~\eqref{eq:column-sum-norm}, this will imply
$\|\advMatrix\|^2=\|\gramMatrix \|\le \sumScale ^2$.

Fix a column indexed by $\fourierKet{v}$ and write $j=\wt(v)$.

\emph{Diagonal part.}
By~\Cref{lem:planted-sum-identities},
\begin{equation}\label{eq:diagonal-bound}
 \begin{aligned}
 \fourierBra{v}\gramMatrix\fourierKet{v}
 &=\sum_{S\in\binom{[\sumSize]}{\sumArity}}
       \|\sumBlock_S\fourierKet{v}\|^2\\
 &\le\binom{\sumSize}{\sumArity}\sumCoeff_j^2
   =(\sumScale-j)_+^2.
 \end{aligned}
\end{equation}

\emph{Off-diagonal part.}
For a fixed certificate $S$, a nonzero term requires a common shift
between $u$ and $v$ on $S$.  Write this shift as $a\in\widehat\sumGroup$:
\begin{equation}\label{eq:common-shift-coordinates}
 u_h=\begin{cases}
       v_h-a,&h\in S,\\
       v_h,&h\notin S.
     \end{cases}
\end{equation}
For an off-diagonal term, $u\ne v$, so $a\ne0$.
If $S\cap\supp(v)=\varnothing$, then
\[
 u_h=-a\ne0\quad(h\in S)
 \quad\Longrightarrow\quad S\subseteq\supp(u).
\]
This contradicts the condition $S\not\subseteq\supp(u)$
in~\eqref{eq:fourier-block}.  Thus we only need certificates that meet
$\supp(v)$.  Choosing one position in this intersection gives
\[
 \#\{S\in\tbinom{[\sumSize]}{\sumArity}:S\cap\supp(v)\ne\varnothing\}
 \le\sum_{h\in\supp(v)}\binom{\sumSize-1}{\sumArity-1}
 =j\binom{\sumSize-1}{\sumArity-1}.
\]
For each such $S$, the condition on $u$ also gives
\[
 S\not\subseteq\supp(u)
 \quad\Longrightarrow\quad
 \exists h\in S:\ u_h=0
 \quad\Longrightarrow\quad
 a\in\{v_h:h\in S\}.
\]
Thus there are at most $\sumArity$ possible shifts.
Each shift determines
one $u$, and its contribution has absolute value at most
$\sumCoeff_j\sumCoeff_{\wt(u)}\le\sumCoeff_j\sumCoeff_0$.
By the triangle inequality,
\begin{align*}
 R_j
 &:=\sum_{u\ne v}|\fourierBra{u}\gramMatrix\fourierKet{v}|\\
 &\le\sumArity j\binom{\sumSize-1}{\sumArity-1}\sumCoeff_j\sumCoeff_0\\
 &=\sumArity j
    \frac{\binom{\sumSize-1}{\sumArity-1}}{\binom{\sumSize}{\sumArity}}
    \sumScale(\sumScale-j)_+\\
 &=\frac{\sumArity^2j}{\sumSize}\sumScale(\sumScale-j)_+,
\end{align*}
where the last equality uses
$\binom{\sumSize-1}{\sumArity-1}/\binom{\sumSize}{\sumArity}
=\sumArity/\sumSize$.

\emph{Combining the two parts.}
If $0\le j<\sumScale$, the assumption
$\sumArity^2\sumScale/\sumSize\le1$ gives $R_j\le j(\sumScale-j)$.
Together with~\eqref{eq:diagonal-bound}, this gives
\begin{align*}
 \sum_u|\fourierBra{u}\gramMatrix\fourierKet{v}|
 &=\fourierBra{v}\gramMatrix\fourierKet{v}+R_j\\
 &\le(\sumScale-j)^2+j(\sumScale-j)\\
 &=\sumScale(\sumScale-j)\le\sumScale^2.
\end{align*}
If $j\ge\sumScale$, then $\sumCoeff_j=0$, and the block action gives
\[
 \advMatrix\fourierKet{v}=0
 \quad\Longrightarrow\quad
 \gramMatrix\fourierKet{v}=0.
\]
Thus every absolute column sum is at most $\sumScale^2$, and
\[
 \|\advMatrix\|^2=\|\gramMatrix\|\le\sumScale^2.
\]
For the lower bound, use~\eqref{eq:published-overlap} and the fact that
the two distribution states have norm one:
\[
 \sumScale=\distBra{\plantedDist}\advMatrix\distKet{\nullDist}
 \le\|\advMatrix\|.
\]
This proves $\|\advMatrix\|=\sumScale$.

\emph{Alignment.}
Equation~\eqref{eq:published-overlap} already gives
$\advMatrix\distKet{\nullDist}=\sumScale\distKet{\plantedDist}$.
For the adjoint relation, expand the squared distance:
\begin{align*}
 &\|\advMatrix^{\adj}\distKet{\plantedDist}
       -\sumScale\distKet{\nullDist}\|^2\\
 &\quad=\|\advMatrix^{\adj}\distKet{\plantedDist}\|^2
       -2\sumScale\operatorname{Re}
         (\distBra{\plantedDist}\advMatrix\distKet{\nullDist})
       +\sumScale^2\\
 &\quad\le\sumScale^2-2\sumScale^2+\sumScale^2=0.
\end{align*}
The squared norm is nonnegative, so it is zero.  Therefore
$\advMatrix^{\adj}\distKet{\plantedDist}=\sumScale\distKet{\nullDist}$,
as required by~\eqref{eq:left-align}.
\end{proof}

\begin{remark}\label{rem:exact-uniform}
The assumption in \Cref{thm:exact-norm} also applies when $\sumArity$
grows with $\sumSize$.
It is equivalent to
$\sumSize ^{1/(\sumArity +1)}\ge \sumArity ^2$.
\end{remark}

The key step is to use the diagonal bound $(\sumScale-j)^2$ and keep
the factor $\sumScale-j$ in the off-diagonal bound.  Their sum is then
at most $\sumScale^2$.  This proves the exact norm for every pair of
parameters satisfying the assumption.

\subsection{The effect of one query}
We now bound the norm of the adversary operator after applying a query
filter.
We make
the dependence on $\sumArity$ explicit in the calculation
of~\cite{ITCS:BelSpa13}.
A bound of the form $O_\sumArity(1)$ would not
tell us how the constant grows with $\sumArity$.
We first use a general
nonincreasing coefficient sequence, then substitute the linear sequence
from~\eqref{eq:alpha-definition}.

\begin{lemma}
  \label[lemma]{lem:filtered-decomposition}
Let $\sumSize \ge 2\sumArity $ and let
\[
 \advMatrix(\sumCoeff)
 =\sum_{S\in\binom{[\sumSize]}{\sumArity}}\ket{S}\otimes\sumBlock_S(\sumCoeff)
\]
be the planted $\sumArity $-SUM matrix formed with a nonnegative nonincreasing sequence
$\sumCoeff =(\sumCoeff _a)_{a\ge0}$, extended by zero past its support.  For every
input position $i$,
\begin{equation}\label{eq:general-filter-decomposition}
 \|\advMatrix (\sumCoeff )\circ\queryDiff _i^{\rm word}\|
 \le 2\bigl(\sqrt{X(\sumCoeff )}+\sqrt{Y(\sumCoeff )}\bigr),
\end{equation}
where
\begin{align}
 X(\sumCoeff )
   &=\max_{a\ge0}\sumCoeff _a^2
       \binom{a+\sumArity -1}{\sumArity -1},\label{eq:general-X}\\
 Y(\sumCoeff )
   &=2\sumArity \binom{\sumSize -1}{\sumArity }
       \max_{a\ge0}|\sumCoeff _a-\sumCoeff _{a+1}|^2.
       \label{eq:general-Y}
\end{align}
\end{lemma}

\begin{proof}
Recall $\fourierProjector_0=\ketbra{\fourierVector_0}{\fourierVector_0}$ and $\fourierProjector_1=\identity-\fourierProjector_0$, and the block definition
\eqref{eq:planted-sum-matrix}.  We first replace the operator by one with the same entries wherever the
query mask is nonzero.
We then separate the terms with coefficient
$\sumCoeff_t$ from those with coefficient $\sumCoeff_t-\sumCoeff_{t+1}$.

Fix $i$.
In every tensor in~\eqref{eq:planted-sum-matrix}, replace the
factor at coordinate $i$ according to
\begin{equation}\label{eq:filter-replacement}
 \fourierProjector _0\longmapsto \fourierProjector _0,\qquad \fourierProjector _1\longmapsto-\fourierProjector _0,
\end{equation}
and call the resulting operator $\advMatrix_i'$.  More explicitly,
\[
 \begin{aligned}
 \sumBlock'_{S,i}(\sumCoeff)
 &=\sqrt{\groupSize}\,\restrictionOp_S^w
   \sum_{U:\,S\not\subseteq U}
      (-1)^{\mathbf1[i\in U]}\sumCoeff_{|U|}
      \fourierProjector_{U\setminus\{i\}},\\
 \advMatrix_i'
 &=\sum_{S\in\binom{[\sumSize]}{\sumArity}}
      \ket{S}\otimes\sumBlock'_{S,i}(\sumCoeff).
 \end{aligned}
\]
For distinct oracle symbols $a,b\in\sumGroup$,
\[
  \matrixel{a}{\fourierProjector_1}{b}
  =\matrixel{a}{\identity-\fourierProjector_0}{b}
 =-\frac1{\groupSize}
 =-\matrixel{a}{\fourierProjector_0}{b}.
\]
Thus the original and replaced factors agree exactly on
the entries retained by the query filter.
That is,
\[
 \advMatrix (\sumCoeff )\circ\queryDiff _i^{\rm word}
   =\advMatrix _i'\circ\queryDiff _i^{\rm word}.
 \]
For any operator $W$ on these row and column label spaces,
let $P_a$ and $Q_a$ project onto the labels whose answer at $i$ is $a$:
\[
 P_a=\sum_{u:\,x_i(u)=a}\ketbra{u}{u},\qquad
 Q_a=\sum_{v:\,x_i(v)=a}\ketbra{v}{v}.
\]
Write $M=\max_a\|P_aWQ_a\|$.
Since the projectors $P_a$ have orthogonal ranges, every input
vector $\ket{\psi}$ satisfies
\[
 \left\|\sum_aP_aWQ_a\ket{\psi}\right\|^2
 =\sum_a\|P_aWQ_a\ket{\psi}\|^2
 \le M^2\sum_a\|Q_a\ket{\psi}\|^2
 =M^2\|\ket{\psi}\|^2.
\]
Thus,%
\[
 \left\|\sum_aP_aWQ_a\right\|
 \le M \le\|W\|.
\]
Subtracting this part gives
\begin{equation}\label{eq:filter-factor-two}
 \|W\circ\queryDiff_i^{\rm word}\|
 =\left\|W-\sum_aP_aWQ_a\right\|
 \le\|W\|+\left\|\sum_aP_aWQ_a\right\|
 \le2\|W\|.
\end{equation}

We will now bound $\|\advMatrix_i'\|$, by using~\Cref{eq:column-sum-norm} in the Fourier basis.
We must allow
for different Fourier states becoming proportional after restriction to
the planted equation.
Write
\[
 \beta_t=\sumCoeff _t-\sumCoeff _{t+1},
 \qquad b=\max_{t\ge0}|\beta_t|.
\]
For $v\in\widehat \sumGroup ^\sumSize $, let $\fourierKet{v}$ be the corresponding Fourier basis state,
write $\fourierSupport =\supp(v)$ and $t=|\fourierSupport |$, and for a certificate $S$ write
\[
 \restrictedKet{S}{v}=\sqrt{\groupSize}\,\restrictionOp_S^w\fourierKet{v}.
\]
This is the unit vector from \Cref{lem:planted-sum-identities}.
We will use the common-shift rule to decide when two such states are
orthogonal.
The replacement keeps only the constant Fourier state at coordinate $i$:
\[
 \fourierProjector_0\fourierKet{v_i}
 =\mathbf1[v_i=0]\fourierKet{0}.
\]
Every term defining $\advMatrix_i'$ therefore vanishes on $\fourierKet{v}$
when $v_i\ne0$.
Assume $v_i=0$, so $i\notin\fourierSupport$.
Only the supports $\fourierSupport$ and $\fourierSupport\cup\{i\}$ can
contribute after the replacement.
Their coefficients are
$\sumCoeff_t$ and $-\sumCoeff_{t+1}$, respectively.
Thus
\[
 \begin{aligned}
 \sumBlock'_{S,i}\fourierKet{v}
 =\left(\sumCoeff_t\mathbf1[S\not\subseteq\fourierSupport]
        -\sumCoeff_{t+1}
          \mathbf1[S\not\subseteq\fourierSupport\cup\{i\}]\right)
          \restrictedKet{S}{v}.
 \end{aligned}
\]
Using $\beta_t=\sumCoeff_t-\sumCoeff_{t+1}$ gives the four cases
\begin{equation}\label{eq:filtered-block-action}
 \sumBlock'_{S,i}\fourierKet{v}=
 \begin{cases}
  \sumCoeff _t\restrictedKet{S}{v},
     &i\in S,\ S\setminus\{i\}\subseteq \fourierSupport ,\\
  \beta_t\restrictedKet{S}{v},
     &i\in S,\ S\setminus\{i\}\not\subseteq \fourierSupport ,\\
  \beta_t\restrictedKet{S}{v},
     &i\notin S,\ S\not\subseteq \fourierSupport ,\\
  0,&i\notin S,\ S\subseteq \fourierSupport .
 \end{cases}
\end{equation}

Let $U_i$ contain the first case and $V_i$ the two cases with coefficient
$\beta_t$.
Then $\advMatrix_i'=U_i+V_i$.  We will prove
$\|U_i\|^2\le X(\sumCoeff)$ and $\|V_i\|^2\le Y(\sumCoeff)$.

We first bound $U_i$.
Fix $S\ni i$.
If $\restrictedKet{S}{u}$ and $\restrictedKet{S}{v}$
have a nonzero inner product, their labels differ by a common shift $a$
as in~\eqref{eq:common-shift-coordinates}.
We only need to consider $u_i=v_i=0$, since \(U_i\) annihilates every other Fourier state.
Thus that shift must be
zero, and $u=v$.
Thus $U_i^{\dagger}U_i$ is diagonal in the character basis.
A certificate contributes to $U_i\fourierKet{v}$ exactly when
\[
 i\in S,\qquad S\setminus\{i\}\subseteq\fourierSupport,
 \qquad |S|=\sumArity.
\]
Since $v_i=0$, we have $i\notin\fourierSupport$.  Thus these certificates
are precisely $S=\{i\}\cup R$ with
$R\subseteq\fourierSupport$ and $|R|=\sumArity-1$.
There are $\binom{t}{\sumArity-1}$ choices, and each contributes
$\sumCoeff_t^2$.  For $v_i=0$, this gives
\[
 \fourierBra{v}U_i^{\adj}U_i\fourierKet{v}
 =\sumCoeff_t^2\binom{t}{\sumArity-1},
\]
where the binomial coefficient is zero when $t<\sumArity-1$.
For $v_i\ne0$, the diagonal entry is zero.  Therefore
\begin{equation}\label{eq:Ui-bound}
 \|U_i\|^2
 \le \max_{t\ge0}\sumCoeff _t^2\binom{t}{\sumArity -1}
 \le \max_{a\ge0}\sumCoeff _a^2\binom{a+\sumArity -1}{\sumArity -1}
 =X(\sumCoeff ).
\end{equation}

Split the row blocks of $V_i$ into $V_i^{\rm in}$ for $i\in S$ and
$V_i^{\rm out}$ for $i\notin S$.  When $i\in S$, both input Fourier vectors still have coordinate zero at
$i$.
A common shift must therefore be zero, exactly as for $U_i$.
Thus $(V_i^{\rm in})^{\adj}V_i^{\rm in}$ is diagonal.
There are at most
$\binom{\sumSize -1}{\sumArity -1}$ certificates containing $i$, each contributing at most
$b^2$, so
\begin{equation}\label{eq:Vi-in-bound}
 \|V_i^{\rm in}\|^2\le\binom{\sumSize -1}{\sumArity -1}b^2.
\end{equation}
When $i\notin S$, the equation $u_i=v_i=0$ does not restrict the shift
on $S$.  Different output states may then have nonzero inner product.
We count how many such states are possible.
Write
$\gramMatrix =(V_i^{\rm out})^{\adj}V_i^{\rm out}$.
Fix a character column $\fourierKet{v}$.
For one
certificate $S\not\ni i$, a character $\fourierKet{u}$ can contribute to the same
column of $\gramMatrix $ only if $u$ agrees with $v$ outside $S$ and is obtained from
$v$ by a common shift on $S$.
Also $S\not\subseteq\supp(u)$, so some
coordinate of $u$ on $S$ is zero.
This determines the common shift from one
of the at most $\sumArity $ character values of $v$ on $S$.
Thus each $S$ contributes
to at most $\sumArity $ entries of the column, each of absolute value at most $b^2$.
Summing over the $\binom{\sumSize-1}{\sumArity}$ possible certificates gives
\[
 \max_v\sum_u|\fourierBra{u}\gramMatrix\fourierKet{v}|
 \le\sum_{\substack{S\in\binom{[\sumSize]}{\sumArity}\\i\notin S}}
       \sumArity b^2
 =\sumArity\binom{\sumSize-1}{\sumArity}b^2.
\]
By~\eqref{eq:column-sum-norm},
\begin{equation}\label{eq:Vi-out-bound}
 \|V_i^{\rm out}\|^2=\|\gramMatrix \|
 \le \sumArity \binom{\sumSize -1}{\sumArity }b^2.
\end{equation}
The two certificate classes occupy disjoint row blocks, so
\[
 V_i^{\adj}V_i=(V_i^{\rm in})^{\adj}V_i^{\rm in}
          +(V_i^{\rm out})^{\adj}V_i^{\rm out}.
\]
Also,
$\binom{\sumSize -1}{\sumArity -1}/\binom{\sumSize -1}{\sumArity }=\sumArity /(\sumSize -\sumArity )\le1$ because $\sumSize \ge2\sumArity $.
Taking norms, and then using $\sumArity +1\le2\sumArity $, gives
\begin{align}
 \|V_i\|^2
 &\le \|V_i^{\rm in}\|^2+\|V_i^{\rm out}\|^2\notag\\
 &\le 2\sumArity \binom{\sumSize -1}{\sumArity }b^2
 =Y(\sumCoeff ).
 \label{eq:Vi-bound}
\end{align}
This gives an explicit count for the two certificate classes
in~\cite[Lemma~10]{ITCS:BelSpa13}.
The count is needed because the
restricted Fourier states are not necessarily orthogonal.

Therefore
\[
 \|\advMatrix _i'\|
 \le\|U_i\|+\|V_i\|
 \le\sqrt{X(\sumCoeff )}+\sqrt{Y(\sumCoeff )}.
\]
Combining this with~\eqref{eq:filter-factor-two} proves
\eqref{eq:general-filter-decomposition}.
\end{proof}

We now evaluate these bounds for the linear coefficients.

\begin{lemma}%
  \label[lemma]{lem:uniform-filter}
Let $\sumSize \ge \sumArity \ge2$, write $\sumScale =\sumSize ^{\sumArity /(\sumArity +1)}$, and suppose
\[
 \frac{\sumArity ^2}{\sumScale }\le\frac12,
 \qquad
 \frac{\sumArity ^2}{\sumSize }\le\frac12.
\]
Then the planted $\sumArity $-SUM matrix satisfies, for every input position $i$,
\[
 \|\advMatrix _{\sumArity ,\sumSize }\circ\queryDiff _i^{\rm word}\|\le9\sqrt{\sumArity }.
\]
In particular, the conclusion holds whenever the assumption
$\sumArity ^2\sumSize ^{-1/(\sumArity +1)}\le1$ of \Cref{thm:exact-norm} holds.
\end{lemma}

\begin{proof}
The assumption $\sumArity ^2/\sumSize \le1/2$ gives $\sumSize \ge2\sumArity ^2\ge2\sumArity $.  Applying
\Cref{lem:filtered-decomposition} to the coefficients
in~\eqref{eq:alpha-definition} gives
\begin{equation}\label{eq:filter-XY}
 \|\advMatrix _{\sumArity ,\sumSize }\circ\queryDiff _i^{\rm word}\|
 \le2(\sqrt X+\sqrt Y),
\end{equation}
where
\begin{align*}
 X&=\max_{a\ge0}
    \sumCoeff _a^2\binom{a+\sumArity -1}{\sumArity -1},\\
 Y&=2\sumArity \binom{\sumSize -1}{\sumArity }
    \max_{a\ge0}|\sumCoeff _a-\sumCoeff _{a+1}|^2,
\end{align*}
and
\[
 \sumCoeff _a=\binom \sumSize \sumArity ^{-1/2}(\sumScale -a)_+.
\]
Since $|\sumCoeff _a-\sumCoeff _{a+1}|\le\binom \sumSize \sumArity ^{-1/2}$,
\begin{equation}\label{eq:Y-bound}
 Y\le 2\sumArity \frac{\binom{\sumSize -1}{\sumArity }}{\binom \sumSize \sumArity }
   =2\sumArity \frac{\sumSize -\sumArity }{\sumSize }\le 2\sumArity .
\end{equation}

It remains to bound $X$.  Only integers $0\le a<\sumScale $ contribute.  Using
\[
 \binom{a+\sumArity -1}{\sumArity -1}
 \le\frac{(a+\sumArity -1)^{\sumArity -1}}{(\sumArity -1)!},
 \qquad
 \binom \sumSize \sumArity \ge\frac{(\sumSize -\sumArity +1)^\sumArity }{\sumArity !},
\]
we get
\begin{equation}\label{eq:X-first}
 X\le
 \sumArity \max_{0\le a<\sumScale }
 \frac{(\sumScale -a)^2(a+\sumArity -1)^{\sumArity -1}}{(\sumSize -\sumArity +1)^\sumArity }.
\end{equation}
Set $R=\sumScale +\sumArity -1$ and $u=a+\sumArity -1$.  Enlarging the range of $u$ to $[0,R]$ can only increase the maximum.
For $0<u<R$, the derivative of $f(u)=(R-u)^2u^{\sumArity -1}$ is
\[
 f'(u)=(R-u)u^{\sumArity -2}\bigl((\sumArity -1)R-(\sumArity +1)u\bigr).
\]
It is positive before $u=(\sumArity -1)R/(\sumArity +1)$ and negative afterward.
The endpoint values are $f(0)=f(R)=0$.  Therefore
\[
 \max_{0\le u\le R}f(u)
 =f\!\left(\frac{(\sumArity-1)R}{\sumArity+1}\right)
 =R^{\sumArity+1}
    \frac{4(\sumArity-1)^{\sumArity-1}}{(\sumArity+1)^{\sumArity+1}}.
\]
Since $\sumScale ^{\sumArity +1}=\sumSize ^\sumArity $,~\eqref{eq:X-first} therefore implies
\begin{align}
  X
  &\le 4\sumArity \frac{(\sumArity -1)^{\sumArity -1}}{(\sumArity +1)^{\sumArity +1}}
    \frac{(\sumScale + \sumArity -1)^{\sumArity + 1}}{(\sumSize - \sumArity +1)^{\sumArity}} \\
  &=4\sumArity \frac{(\sumArity -1)^{\sumArity -1}}{(\sumArity +1)^{\sumArity +1}}
    \left(1+\frac{\sumArity -1}{\sumScale }\right)^{\sumArity +1}
    \left(1-\frac{\sumArity -1}{\sumSize }\right)^{-\sumArity }.\label{eq:X-product}
\end{align}
We bound the two correction factors separately.
Since
$\log(1+x)\le x$ for $x\ge0$,
\[
 \log\left(1+\frac{\sumArity -1}{\sumScale }\right)^{\sumArity +1}
 \le\frac{\sumArity ^2-1}{\sumScale }\le\frac12\le1.
\]
Thus the first correction factor is at most $e$.
For the second, write
$x=(\sumArity -1)/\sumSize \le1/2$ and use $-\log(1-x)\le2x$:
\[
 \log\left(1-\frac{\sumArity -1}{\sumSize }\right)^{-\sumArity }
 \le\frac{2\sumArity (\sumArity -1)}\sumSize \le1.
\]
It is also at most $e$.
Finally,
\[
 \sumArity \frac{(\sumArity -1)^{\sumArity -1}}{(\sumArity +1)^{\sumArity +1}}
 =\frac{\sumArity }{(\sumArity +1)^2}
       \left(\frac{\sumArity -1}{\sumArity +1}\right)^{\sumArity -1}
 \le\frac{\sumArity }{(\sumArity +1)^2}\le\frac1\sumArity .
\]
Substitution into~\eqref{eq:X-product} gives
\begin{equation}\label{eq:X-bound}
 X\le\frac{4e^2}{\sumArity }.
\end{equation}
Combining~\eqref{eq:filter-XY}, \eqref{eq:Y-bound}, and
\eqref{eq:X-bound},
\[
 \|\advMatrix _{\sumArity ,\sumSize }\circ\queryDiff _i^{\rm word}\|
 \le \frac{4e}{\sqrt{\sumArity }}+2\sqrt{2\sumArity }
 \le9\sqrt{\sumArity }.
\]
For the last inequality, factor by $\sqrt{\sumArity }$ and use
$4e/\sumArity +2\sqrt2\le2e+2\sqrt2<9$, since $\sumArity \ge2$.

For the final statement, $\sumArity ^2\sumSize ^{-1/(\sumArity +1)}\le1$ implies
$\sumSize \ge \sumArity ^{2(\sumArity +1)}$.
Thus
\[
 \frac{\sumArity ^2}{\sumScale }\le \sumArity ^{2-2\sumArity }\le\frac14,
 \qquad
 \frac{\sumArity ^2}{\sumSize }\le \sumArity ^{-2\sumArity }\le\frac1{16},
\]
so the assumptions of the lemma hold.
\end{proof}

\subsection{From word filters to bit filters}

The outer labels will be elements of $\F_2^\outerBits $, while each bit is encoded
separately.
We therefore need the same bound for a bit filter.
Different bits imply different words.  However, this observation alone
does not prove the norm inequality: setting some entries of a signed
matrix to zero can increase its norm.  We use the following phase-operator
identity.

\begin{lemma}
  \label[lemma]{lem:bit-filter}
Let $\advMatrix $ be any rectangular matrix whose row and column labels contain
words $x_i,y_i\in\F_2^\outerBits $.  Then, for every bit position $t$,
\[
 \|\advMatrix \circ\queryDiff _{i,t}^{\rm bit}\|
 \le
 \|\advMatrix \circ\queryDiff _i^{\rm word}\|.
\]
\end{lemma}

\begin{proof}
Let $A=\advMatrix\circ\queryDiff_i^{\rm word}$.  Define the diagonal
phase operators on the row and column registers by
\[
 R=\sum_x(-1)^{x_i(t)}\ketbra{x}{x},
 \qquad
 S=\sum_y(-1)^{y_i(t)}\ketbra{y}{y}.
\]
Their action gives
\[
 \matrixel{x}{RAS}{y}
 =(-1)^{x_i(t)+y_i(t)}\matrixel{x}{A}{y}.
\]
Therefore,
\begin{align*}
 \matrixel{x}{\tfrac12(A-RAS)}{y}
 &=\frac{1-(-1)^{x_i(t)+y_i(t)}}2\matrixel{x}{A}{y}\\
 &=\mathbf1[x_i(t)\ne y_i(t)]\matrixel{x}{A}{y}\\
 &=\matrixel{x}{\advMatrix\circ\queryDiff_{i,t}^{\rm bit}}{y}.
\end{align*}
The last equality uses the fact that different bits imply different words,
so the word filter has already retained those entries.
Thus
\[
 \advMatrix\circ\queryDiff_{i,t}^{\rm bit}=\tfrac12(A-RAS).
\]
The matrices $R,S$ are unitary, so
\[
  \|\advMatrix \circ\queryDiff _{i,t}^{\rm bit}\|
  \le\frac12(\|A\|+\|RAS\|)=\|A\|.
\]
\end{proof}

\section{Composition for the planted distributions}\label{sec:composition}

We use the matrix composition from~\Cref{lem:completed-upper}.
The remaining step is to check that the
construction preserves the overlap for our chosen distributions.
The planted and null distributions can assign positive probability to
the same oracle string.
Thus the encoded bit does not need to be a function
of the string seen by the algorithm, and the function-composition
statement alone does not give the required distributional bound.
The alignment equations from~\Cref{thm:exact-norm} give the overlap
calculation below.
We then apply it to the two planted-sum problems.

\subsection{The encoded distributions and their operator}

Let $\bitDist_0,\bitDist_1$ be distributions on labelled inner oracle inputs.
Set
\[
 \mathcal H_b=\labelSpace{\supp(\bitDist_b)},\qquad
 \distKet{b}=\distKet{\bitDist_b}
 \quad(b\in\{0,1\}).
\]
Let $\innerAdv:\mathcal H_0\to\mathcal H_1$ satisfy
\begin{equation}\label{eq:inner-alignment-assumption}
 \|\innerAdv\|=\innerNorm,\qquad
 \innerAdv\distKet{0}=\innerNorm\distKet{1},\qquad
 \innerAdv^{\adj}\distKet{1}=\innerNorm\distKet{0}.
\end{equation}
For all pairs of outer bits, use the completed operators
\[
 \innerAdv^{1,0}=\innerAdv,\quad
 \innerAdv^{0,1}=\innerAdv^{\adj},\quad
 \innerAdv^{b,b}=\innerNorm\identity_{\mathcal H_b}
 \quad(b\in\{0,1\}).
\]
Thus $\innerAdv^{b,c}:\mathcal H_c\to\mathcal H_b$.
The identity operators define adversary weights for equal outer bits.
They do not mean that two independently sampled inner inputs are equal.

Let $\plantedDist_F,\nullDist_F$ be distributions on outer input labels, and let
$\outerAdv:\labelSpace{\supp(\nullDist_F)}
\to\labelSpace{\supp(\plantedDist_F)}$.
Each label $x$ carries a bit string $(x_1,\ldots,x_{\numBlocks})$;
the label may contain additional information.
To sample $\widetilde\plantedDist$, first sample $x\leftarrow\plantedDist_F$
and then independently sample $u_p\leftarrow\bitDist_{x_p}$ for
$p\in[\numBlocks]$.  Thus
\[
 \widetilde\plantedDist[x,u_1,\ldots,u_{\numBlocks}]
 =\plantedDist_F[x]\prod_{p=1}^{\numBlocks}\bitDist_{x_p}[u_p].
\]
Define $\widetilde\nullDist$ similarly from $\nullDist_F$.
We keep all labels in the analysis.  The algorithm sees only the
inner oracle strings.  A physical query $(p,j)$ reads position $j$ in
inner block $p$, and may be in superposition over both indices.

Using the label registers of~\Cref{lem:completed-upper}, define
\begin{equation}\label{eq:composed-matrix}
 \composedAdv
 =\sum_{x,y}\matrixel{x}{\outerAdv}{y}\ketbra{x}{y}
       \otimes\bigotimes_{p=1}^{\numBlocks}\innerAdv^{x_p,y_p}.
\end{equation}
Here and below, $x$ ranges over $\supp(\plantedDist_F)$ and $y$ over
$\supp(\nullDist_F)$.
The factor $\ketbra{x}{y}$ acts on the outer-label register; the other
factors act on the inner registers.  To use fixed register spaces,
embed $\mathcal H_0$ and $\mathcal H_1$ as orthogonal subspaces of
$\mathcal H_0\oplus\mathcal H_1$ and extend all operators by zero.
The operator $\composedAdv$ is zero outside the valid encoded-label spaces.

The outer bit filter and the physical query filter compare different data:
\[
 \matrixel{x}{\queryDiff_p^{\rm bit}}{y}=\mathbf1[x_p\ne y_p],
 \qquad
 \matrixel{x,u}{\queryDiff_{p,j}}{y,v}
 =\mathbf1[u_p(j)\ne v_p(j)],
\]
where $u=(u_1,\ldots,u_{\numBlocks})$ and $u_p(j)$ denotes the oracle
answer specified by the inner label $u_p$ at position $j$.

\begin{lemma}%
  \label[lemma]{lem:composition}
The operator in~\eqref{eq:composed-matrix} satisfies
\begin{align}
 \|\composedAdv\|&=\|\outerAdv\|\innerNorm^{\numBlocks},
       \label{eq:comp-norm}\\
 \distBra{\widetilde\plantedDist}\composedAdv\distKet{\widetilde\nullDist}
 &=\bigl(\distBra{\plantedDist_F}\outerAdv\distKet{\nullDist_F}\bigr)
       \innerNorm^{\numBlocks}.
       \label{eq:comp-overlap}
\end{align}
For every physical query $(p,j)$,
\begin{equation}\label{eq:comp-filter}
 \|\composedAdv\circ\queryDiff_{p,j}\|
 \le\|\outerAdv\circ\queryDiff_p^{\rm bit}\|
       \innerNorm^{\numBlocks-1}\|\innerAdv\circ\queryDiff_j\|.
\end{equation}
\end{lemma}

\begin{proof}
The norm identity~\eqref{eq:comp-norm} and the query-filter bound
\eqref{eq:comp-filter} follow from~\Cref{lem:completed-upper} with
$D_p=\innerAdv$ and $\rho_p=\innerNorm$ for every $p$.
The inner filter $\queryDiff_{p,j}^{\rm in}$ in that lemma is
$\queryDiff_j$ here.
We only need to calculate the distributional overlap.

The alignment assumptions, including the identity cases, give
\begin{equation}\label{eq:all-block-alignment}
 \innerAdv^{b,c}\distKet{c}=\innerNorm\distKet{b}
 \quad\text{for every }b,c\in\{0,1\}.
\end{equation}
For each outer input label $x$, write
\[
 \ket{\eta_x}=\bigotimes_{p=1}^{\numBlocks}\distKet{x_p}.
\]
The square-root states of the encoded distributions are
\[
 \distKet{\widetilde\plantedDist}
 =\sum_x\sqrt{\plantedDist_F[x]}\ket{x}\otimes\ket{\eta_x},
 \qquad
 \distKet{\widetilde\nullDist}
 =\sum_y\sqrt{\nullDist_F[y]}\ket{y}\otimes\ket{\eta_y}.
\]
Since each $\distKet{b}$ is normalized,~\eqref{eq:all-block-alignment}
implies
$\distBra{b}\innerAdv^{b,c}\distKet{c}=\innerNorm$.
Expanding the overlap therefore gives
\begin{align*}
 &\distBra{\widetilde\plantedDist}\composedAdv\distKet{\widetilde\nullDist}\\
 &\quad=\sum_{x,y}\sqrt{\plantedDist_F[x]\nullDist_F[y]}
       \matrixel{x}{\outerAdv}{y}
       \prod_{p=1}^{\numBlocks}
          \distBra{x_p}\innerAdv^{x_p,y_p}\distKet{y_p}\\
 &\quad=\innerNorm^{\numBlocks}
       \sum_{x,y}\sqrt{\plantedDist_F[x]\nullDist_F[y]}
          \matrixel{x}{\outerAdv}{y}\\
 &\quad=\innerNorm^{\numBlocks}
       \distBra{\plantedDist_F}\outerAdv\distKet{\nullDist_F}.
\end{align*}
This proves~\eqref{eq:comp-overlap}.
\end{proof}

\subsection{Applying composition to two planted sum problems}

We first use the direct encoding: a uniform block represents zero and a
planted block represents one.
The next theorem gives a query lower bound for distinguishing the two
encoded distributions.

\begin{theorem}
\label{thm:single-sum-composition}
Let $\innerSize \ge \innerArity \ge2$ and $\outerSize \ge\outerArity \ge2$ satisfy
\begin{equation}\label{eq:composition-conditions}
 \innerArity ^2\innerSize ^{-1/(\innerArity +1)}\le1,\qquad
 \outerArity ^2\outerSize ^{-1/(\outerArity +1)}\le1.
\end{equation}
Fix $\innerBits ,\outerBits \ge1$ and $z\in\F_2^\outerBits $.  The inner distributions are
\[
 \bitDist _0=\nullDist _{\innerArity ,\innerSize },\qquad \bitDist _1=\plantedDist _{\innerArity ,\innerSize }^{0}
\]
on length-$\innerSize $ strings over $\F_2^\innerBits $, with hidden labels in the planted case.
The outer distributions are $\plantedDist _{\outerArity ,\outerSize }^{z}$ and $\nullDist _{\outerArity ,\outerSize }$ on
length-$\outerSize $ strings over $\F_2^\outerBits $.  Write each outer word as $\outerBits $ bits, and
independently replace each bit $b$ by an inner sample from $\bitDist _b$.
If a quantum algorithm accepts the encoded planted and null distributions
with average probabilities $\acceptProb _\plantedDist ,\acceptProb _\nullDist $, its worst-case query count is at least
\[
 \Omega\!\left((1-\outputOverlap (\acceptProb _\plantedDist ,\acceptProb _\nullDist ))
 \frac{\outerSize ^{\outerArity /(\outerArity +1)}\innerSize ^{\innerArity /(\innerArity +1)}}{\sqrt{\innerArity \outerArity }}\right).
\]
\end{theorem}

\begin{proof}
There are $\numBlocks =\outerSize \outerBits $ inner blocks, indexed by outer bit positions $p=(i,t)$.
Take $\innerAdv =\advMatrix _{\innerArity ,\innerSize }$ with rows distributed as $\bitDist _1$ and columns as
$\bitDist _0$, and take $\outerAdv =\advMatrix _{\outerArity ,\outerSize }$ for the outer target $z$.  Write
\[
 \innerNorm =\innerSize ^{\innerArity /(\innerArity +1)},\qquad \lambda=\outerSize ^{\outerArity /(\outerArity +1)}.
\]
By \Cref{thm:exact-norm},
\[
 \|\innerAdv \|=\innerNorm ,\quad \innerAdv \distKet{0}=\innerNorm \distKet{1},\quad
 \innerAdv ^{\adj}\distKet{1}=\innerNorm \distKet{0},
\]
and
\[
 \|\outerAdv \|=\lambda,\qquad
 \distBra{\plantedDist _{\outerArity ,\outerSize }^{z}}\outerAdv \distKet{\nullDist _{\outerArity ,\outerSize }}=\lambda.
\]
By \Cref{lem:uniform-filter,lem:bit-filter}, for every relevant coordinate,
\[
 \|\innerAdv \circ\queryDiff _j\|\le9\sqrt \innerArity ,\qquad
 \|\outerAdv \circ\queryDiff _p^{\rm bit}\|\le9\sqrt\outerArity .
\]
\Cref{lem:composition} therefore gives
\begin{align*}
 \|\composedAdv \|&=\lambda\innerNorm ^\numBlocks ,\\
 \distBra{\widetilde \plantedDist}\composedAdv \distKet{\widetilde \nullDist}
       &=\lambda\innerNorm ^\numBlocks ,\\
 \max_{p,j}\|\composedAdv \circ\queryDiff _{p,j}\|
       &\le(9\sqrt\outerArity )(9\sqrt \innerArity )\innerNorm ^{\numBlocks -1}
        =81\sqrt{\innerArity \outerArity }\,\innerNorm ^{\numBlocks -1}.
\end{align*}
If $\acceptProb _\plantedDist =\acceptProb _\nullDist $, the stated lower bound is zero and the result follows directly.  Otherwise \Cref{thm:average-adv} yields
\begin{align*}
 \queryBudget
 &\ge\frac{\lambda\innerNorm ^\numBlocks -\outputOverlap (\acceptProb _\plantedDist ,\acceptProb _\nullDist )\lambda\innerNorm ^\numBlocks }
              {162\sqrt{\innerArity \outerArity }\,\innerNorm ^{\numBlocks -1}}\\
 &=\frac{1-\outputOverlap (\acceptProb _\plantedDist ,\acceptProb _\nullDist )}{162}
      \frac{\lambda\innerNorm }{\sqrt{\innerArity \outerArity }}.
\end{align*}
Substituting $\lambda$ and $\innerNorm $ proves the theorem.
\end{proof}

The product of the two query scales comes from the cancellation
\[
 \frac{\lambda\innerNorm^{\numBlocks}}
      {\sqrt{\innerArity\outerArity}\,\innerNorm^{\numBlocks-1}}
 =\frac{\lambda\innerNorm}{\sqrt{\innerArity\outerArity}}.
\]
One physical query acts on one inner block.  The other $\numBlocks-1$
blocks each contribute a factor $\innerNorm$ to the filtered-norm bound.
These factors cancel, leaving one inner factor and one outer factor.
The number $\numBlocks=\outerSize\outerBits$ therefore disappears from
the final ratio.  The proof allows arbitrary superpositions of queries
across all blocks.

\subsection{The paired encoding for perfect agreement}

For the perfect-agreement protocol we will use two blocks for each bit.
The following calculation shows that this change preserves the inner
adversary ratio.

\begin{lemma}%
  \label[lemma]{lem:paired-block}
Assume~\eqref{eq:inner-alignment-assumption}, and define
\[
 \widehat\bitDist _0=\bitDist _1\otimes\bitDist _0,\qquad
 \widehat\bitDist _1=\bitDist _0\otimes\bitDist _1,\qquad
 \widehat \innerAdv =\innerAdv ^{\adj}\otimes \innerAdv .
\]
Then $\widehat \innerAdv $ has rows indexed by $\widehat\bitDist _1$ and columns by
$\widehat\bitDist _0$, and
\begin{align}
 \|\widehat \innerAdv \|&=\innerNorm ^2,\label{eq:paired-norm}\\
 \widehat \innerAdv \distKet{\widehat\bitDist _0}&=\innerNorm ^2\distKet{\widehat\bitDist _1},
 \qquad
 \widehat \innerAdv ^{\adj}\distKet{\widehat\bitDist _1}=\innerNorm ^2\distKet{\widehat\bitDist _0}.
 \label{eq:paired-align}
\end{align}
Let $\queryDiff_{h,j}$ be the filter for a query to entry $j$ in
half $h\in\{1,2\}$ of the pair.  Then
\begin{equation}\label{eq:paired-filter}
 \|\widehat\innerAdv\circ\queryDiff_{h,j}\|
 \le\innerNorm\|\innerAdv\circ\queryDiff_j\|.
\end{equation}
\end{lemma}

\begin{proof}
The tensor-product norm identity gives
$\|\widehat \innerAdv \|=\|\innerAdv ^{\adj}\|\|\innerAdv \|=\innerNorm ^2$.  The alignment follows from
\begin{align*}
 \widehat \innerAdv \distKet{\widehat\bitDist _0}
 &=(\innerAdv ^{\adj}\otimes \innerAdv )(\distKet{1}\otimes\distKet{0})\\
 &=(\innerAdv ^{\adj}\distKet{1})\otimes(\innerAdv \distKet{0})\\
 &=(\innerNorm \distKet{0})\otimes(\innerNorm \distKet{1})
   =\innerNorm ^2\distKet{\widehat\bitDist _1}.
\end{align*}
The adjoint relation follows in the same way:
\begin{align*}
 \widehat\innerAdv^{\adj}\distKet{\widehat\bitDist_1}
 &=(\innerAdv\otimes\innerAdv^{\adj})
       (\distKet{0}\otimes\distKet{1})\\
 &=(\innerAdv\distKet{0})\otimes
       (\innerAdv^{\adj}\distKet{1})\\
 &=\innerNorm^2\distKet{\widehat\bitDist_0}.
\end{align*}
For the two possible query positions, the filtered operators are
\[
 \begin{aligned}
 \widehat\innerAdv\circ\queryDiff_{1,j}
   &=(\innerAdv\circ\queryDiff_j)^{\adj}\otimes\innerAdv,\\
 \widehat\innerAdv\circ\queryDiff_{2,j}
   &=\innerAdv^{\adj}\otimes(\innerAdv\circ\queryDiff_j).
 \end{aligned}
\]
The first equality uses
$(\innerAdv\circ\queryDiff_j)^{\adj}
=\innerAdv^{\adj}\circ\queryDiff_j^{\mathsf T}$,
since the difference mask is real.
Using the tensor-product norm identity gives, for both $h=1$ and $h=2$,
\[
 \|\widehat\innerAdv\circ\queryDiff_{h,j}\|
 =\|\innerAdv\|\|\innerAdv\circ\queryDiff_j\|
 =\innerNorm\|\innerAdv\circ\queryDiff_j\|.
\]
\end{proof}

\begin{theorem}[Paired planted-sum composition]
\label{thm:paired-sum-composition}
Under the parameter assumptions of \Cref{thm:single-sum-composition},
replace the inner encoding by
\[
 \widehat\bitDist _0=\plantedDist _{\innerArity ,\innerSize }^{0}\otimes \nullDist _{\innerArity ,\innerSize },\qquad
 \widehat\bitDist _1=\nullDist _{\innerArity ,\innerSize }\otimes \plantedDist _{\innerArity ,\innerSize }^{0}.
\]
The outer distributions are again $\plantedDist _{\outerArity ,\outerSize }^{z}$ and $\nullDist _{\outerArity ,\outerSize }$, with
independent paired blocks for their $\numBlocks =\outerSize \outerBits $ bits.  Every quantum algorithm
with average acceptance probabilities $\acceptProb _\plantedDist ,\acceptProb _\nullDist $ makes at least
\[
 \Omega\!\left((1-\outputOverlap (\acceptProb _\plantedDist ,\acceptProb _\nullDist ))
 \frac{\outerSize ^{\outerArity /(\outerArity +1)}\innerSize ^{\innerArity /(\innerArity +1)}}{\sqrt{\innerArity \outerArity }}\right)
\]
queries in the worst case.  A physical query reads one entry of one half
of one pair.
\end{theorem}

\begin{proof}
Use $\innerAdv,\outerAdv,\innerNorm,\lambda$ from the single-block proof.
\Cref{lem:paired-block} gives exact alignment for $\widehat \innerAdv $ with
$\widehat\innerNorm =\innerNorm ^2$, and its local filtered norm is at most
$9\sqrt \innerArity \,\innerNorm $.  The outer bit-filtered norm is at most
$9\sqrt\outerArity $.  Applying \Cref{lem:composition} to $\outerAdv $ and $\widehat \innerAdv $
gives the following bounds, where $(p,h,j)$ denotes entry $j$ in half
$h$ of the pair for outer bit $p$:
\[
 \|\composedAdv \|=\distBra{\widetilde \plantedDist}\composedAdv \distKet{\widetilde \nullDist}
       =\lambda\widehat\innerNorm ^\numBlocks ,
 \qquad
 \max_{p,h,j}\|\composedAdv \circ\queryDiff _{p,h,j}\|
       \le81\sqrt{\innerArity \outerArity }\,\widehat\innerNorm ^{\numBlocks -1}\innerNorm .
\]
For $\acceptProb _\plantedDist \ne \acceptProb _\nullDist $, \Cref{thm:average-adv} yields
\begin{align*}
 \queryBudget
 &\ge\frac{(1-\outputOverlap (\acceptProb _\plantedDist ,\acceptProb _\nullDist ))\lambda\widehat\innerNorm ^\numBlocks }
              {162\sqrt{\innerArity \outerArity }\,\widehat\innerNorm ^{\numBlocks -1}\innerNorm }\\
 &=\frac{1-\outputOverlap (\acceptProb _\plantedDist ,\acceptProb _\nullDist )}{162}
       \frac{\lambda}{\sqrt{\innerArity \outerArity }}\frac{\innerNorm ^2}{\innerNorm }
  =\frac{1-\outputOverlap (\acceptProb _\plantedDist ,\acceptProb _\nullDist )}{162}
       \frac{\lambda\innerNorm }{\sqrt{\innerArity \outerArity }}.
\end{align*}
This is the same bound as for one block.  The case $\acceptProb _\plantedDist =\acceptProb _\nullDist $ is again
immediate.
\end{proof}

\section{The key agreement protocol}
\label{sec:imperfect-protocol}

We first use the single-block composition to define a classical
random-oracle protocol.
Fix integers $\innerArity ,\outerArity \ge2$ and
$\securityParam \ge\max\{\innerArity ,\outerArity \}$.
Set
\[
 \outerBits =\left\lceil(\outerArity +3)\log_2\securityParam \right\rceil,
 \qquad
 \innerBits =\left\lceil(\innerArity +4)\log_2\securityParam \right\rceil.
\]
The parties have oracle access to a uniformly random function
\[
 \randomOracle :[\securityParam ]\times[\outerBits ]\times[\securityParam ]
   \longrightarrow\F_2^\innerBits .
 \]
 For each \((i,t) \in [\securityParam ]\times[\outerBits ]\), define the oracle block
 \[
\randomOracle_{i,t} := \left(\randomOracle(i,t,1),\dots,\randomOracle(i,t,\securityParam)\right) \in (\F_2^\innerBits)^{\securityParam}.
\]
The index \(i\) identifies an outer label, \(t\) identifies a bit of that label, and \(j \in [\securityParam]\) identifies an entry within the block.
This way, the oracle consists of \(\securityParam\outerBits\) different blocks, each with \(\securityParam\) entries.
Because $\randomOracle $ is a random function, its values in these blocks
are independent.
Here $\securityParam $ plays both length roles from the composition
theorem: $\outerSize =\innerSize =\securityParam $.
In the following, Alice chooses \(\securityParam\) random \(\outerBits\)-bit strings, which we call outer labels.
Alice will encode the $\securityParam \outerBits $ bits of her $\securityParam$ labels each in one block.

A query at \((i,t,j)\) returns the \(\innerBits\)-bit value \(\randomOracle(i,t,j)\).
A full classical read of the block \((i,t)\) consists of \(\securityParam\) queries, for all possible values of \(j\in[\securityParam]\).
Quantum queries can be made in superposition over all three indices.

\subsection{Preparing one masked block}

Fix \((i,t) \in [\securityParam ]\times[\outerBits ]\).
For simplicity, we omit dependence in \((i,t)\) and write, for \(j\in[\securityParam]\),  $\randomOracle (j)=\randomOracle (i,t,j)$.
Alice chooses a public mask for this block, in either null or planted mode.
In both cases she
publishes masks
$\publicMask =(\publicMask _1,\ldots,\publicMask _\securityParam )\in(\F_2^\innerBits )^\securityParam $.
The vector \(\publicMask\) is the mask for block \((i,t)\), where we again omit dependence in \((i,t)\).

\begin{itemize}[leftmargin=2em]
\item In null mode, she samples every $\publicMask _j$ independently and uniformly
      and makes no query.
\item In planted mode, she samples $S\in\binom{[\securityParam ]}\innerArity $ uniformly, chooses a pivot
      $p\in S$, samples $\publicMask _j$ uniformly for all $j\ne p$, queries $\randomOracle (j)$
      for every $j\in S$, and sets
      \[
        \publicMask _p
        =\randomOracle (p)+\sum_{j\in S\setminus\{p\}}
                     \bigl(\randomOracle (j)+\publicMask _j\bigr).
      \]
\end{itemize}
Define the masked block
\[
  Y:= (Y_1,\dots,Y_\securityParam), \quad\quad Y_j:=\randomOracle (j)+\publicMask _j.
\]
Define its relation predicate by
\[
 \relationTest (Y)=1
 \quad\Longleftrightarrow\quad
 \exists S\in\binom{[\securityParam ]}\innerArity :
       \sum_{j\in S}Y_j=0.
\]

\begin{lemma}
  \label[lemma]{lem:mask-distribution}
In null mode, the pair $(\publicMask ,Y)$ is distributed as an independent uniform
mask and a sample from $\nullDist _{\innerArity ,\securityParam }$.
In planted mode, \((\publicMask,S,Y)\) is distributed as an
independent uniform mask \(\publicMask\) and a planted sample $(S,Y) \leftarrow \plantedDist _{\innerArity ,\securityParam }^{0}$.
Also, quantum query access to $\randomOracle $ given $\publicMask $ is
exactly equivalent, one query for one query, to access to $Y$ given $\publicMask $.
\end{lemma}

\begin{proof}
In null mode, the map $(\randomOracle ,\publicMask )\mapsto(Y,\publicMask )$ with $Y=\randomOracle +\publicMask $ is a bijection,
so $Y$ and $\publicMask $ are independent and uniform.

Fix $S$ and $p$ in planted mode.  The construction makes
\[
  Y_p=\sum_{j\in S\setminus\{p\}}Y_j,
\]
and thus $\sum_{j\in S}Y_j=0$.  Consider the map from the initially uniform
variables
\[
  \bigl(\randomOracle ,(\publicMask _j)_{j\ne p}\bigr)
\]
to $(Y,\publicMask )$.  Its image is the set of pairs for which
$\sum_{j\in S}Y_j=0$.  Conversely, given any such pair, the unique preimage is
$\randomOracle =Y+\publicMask $, and the defining equation for $\publicMask _p$ holds because the sum of
the $Y_j$ on $S$ is zero.  The map is therefore a bijection.  It follows that
$Y$ is uniform subject to the planted equation and that $\publicMask $ is uniform and
independent of $Y$.  Averaging over uniform $S$ gives $\plantedDist _{\innerArity ,\securityParam }^{0}$.

Finally, with $\publicMask $ known, a query to $\randomOracle $ can be converted to a
query to $Y$ by adding the known value $\publicMask _j$ to the answer register, and
vice versa.
This conversion is coherent and uses no oracle query.
\end{proof}

For any fixed $\innerArity $-subset of a uniform block, the sum is uniform in
$\F_2^\innerBits $ and equals zero with probability $2^{-\innerBits }$.
Summing over the
$\binom \securityParam \innerArity $ possible subsets gives
\begin{equation}\label{eq:false-positive}
 \Prb_{Y\leftarrow \nullDist _{\innerArity ,\securityParam }}
   [\relationTest (Y)=1]
 \le\frac{\binom \securityParam \innerArity }{2^\innerBits }
 \le\frac{\securityParam ^\innerArity }{\securityParam ^{\innerArity +4}}=\securityParam ^{-4}.
\end{equation}

\subsection{Single-block encoding of one bit}

To encode $b\in\{0,1\}$, Alice prepares one block by
\begin{equation}\label{eq:single-encoding}
 b=0:\quad Y\leftarrow \nullDist _{\innerArity ,\securityParam },
 \qquad
 b=1:\quad Y\leftarrow \plantedDist _{\innerArity ,\securityParam }^{0},
\end{equation}
using the masking procedure above.
She makes no query for a zero bit and
exactly $\innerArity $ queries for a one bit.
Bob reads the block and decodes
\begin{equation}\label{eq:single-decoder}
 \bitDecoder (Y)=\relationTest (Y).
\end{equation}
A one bit is always decoded correctly.
By~\eqref{eq:false-positive}, a zero
bit is decoded incorrectly with probability at most $\securityParam ^{-4}$.

\subsection{The protocol and its correctness}

The protocol is defined as follows.

\begin{enumerate}[leftmargin=2.2em,label=\textbf{\arabic*.}]
\item Alice samples independent uniform labels
$a_1,\ldots,a_\securityParam \in\F_2^\outerBits $.
For every $(i,t)\in[\securityParam ]\times[\outerBits ]$, she
      encodes the bit $a_i(t)$ in block $(i,t)$ according
      to~\eqref{eq:single-encoding} and obtains $\publicMask_{i,t} := (\publicMask_{i,t,j})_{j\in[\securityParam]} $ .
      She sends the entire mask table $m:=(\publicMask_{i,t})_{(i,t)\in[\securityParam]\times[\outerBits]} $ to
      Bob.
      \item Bob samples a uniform $\outerArity $-set $\sharedKey \in\binom{[\securityParam ]}\outerArity $.
      For every
      $i\in \sharedKey $ and $t\in[\outerBits ]$, Bob queries all \(\securityParam\) entries of \(\randomOracle_{i,t}\), computes \(Y_{i,t} = (Y_{i,t,j})_{j\in[\securityParam]}\), \(Y_{i,t,j} = \randomOracle(i,t,j) + m_{i,t,j}\)
      and sets \(\widehat a_i(t) = \bitDecoder(Y_{i,t})\).
      He sends
      \[
        z=\sum_{i\in \sharedKey }\widehat a_i\in\F_2^\outerBits
      \]
      to Alice and outputs $\sharedKey $.
      Upon receiving $z$, Alice searches, using free
      local computation, for the $\outerArity $-subsets $J\subseteq[\securityParam ]$ satisfying
      $\sum_{j\in J}a_j=z$.
      If there is exactly one, she outputs it, otherwise
      she aborts.
\end{enumerate}

\begin{proposition}%
\label[proposition]{prop:imperfect-correctness}
The probability that both parties output the same key is at least
\[
  1-\outerArity  \outerBits \securityParam ^{-4}-\securityParam ^{-3}.
\]
Alice makes at most $\innerArity \securityParam \outerBits $ oracle queries and Bob makes exactly $\outerArity  \outerBits \securityParam $
queries.
Alice's expected number of queries is $\innerArity \securityParam \outerBits /2$.
Thus the total
worst-case query complexity is
$O((\innerArity +\outerArity )\outerArity  \securityParam \log \securityParam )$.
\end{proposition}

\begin{proof}
By~\eqref{eq:false-positive} and a union bound over the at most $\outerArity  \outerBits $
uniform blocks read by Bob, the probability that any decoded bit is
incorrect is at most
\[
  \outerArity  \outerBits \securityParam ^{-4}.
\]
If no such error occurs, then $\widehat a_i=a_i$ for all $i\in \sharedKey $ and thus
$\sharedKey $ is a solution to Alice's reconstruction equation.
We bound the second bad event using the original labels, without
conditioning on successful inner decoding.
Condition only on $\sharedKey $ and
$z_{\rm ideal}=\sum_{i\in \sharedKey }a_i$, and write this target as $z$ for the
following calculation.
For any fixed $J\ne \sharedKey $ of size
$\outerArity $, choose an index in $J\setminus \sharedKey $.
Its label remains uniform after
conditioning on the sum over $\sharedKey $.
Conditioning on every other label, the
equation $\sum_{j\in J}a_j=z$ therefore holds with probability exactly
$2^{-\outerBits }$.
A union bound gives
\[
 \Prb[\exists J\ne \sharedKey :\sum_{j\in J}a_j=z]
 \le\frac{\binom \securityParam \outerArity }{2^\outerBits }
 \le\frac{\securityParam ^\outerArity }{\securityParam ^{\outerArity +3}}=\securityParam ^{-3}.
\]
The two bad events are not necessarily independent, so we only add their probabilities.
Outside these two bad events, Alice's unique solution is $\sharedKey $, and in this case the parties
output the same key.
This proves the correctness bound.

Alice makes $\innerArity $ queries for each bit equal to one among the $\securityParam \outerBits $ independent uniform label
bits, thus at most $\innerArity \securityParam \outerBits $ queries and $\innerArity \securityParam \outerBits /2$ in expectation.
Bob reads one
length-$\securityParam $ block for each of the $\outerArity  \outerBits $ selected bits, and reading one block requires exactly N queries.
Finally, using
$\outerBits =O(\outerArity \log \securityParam )$ proves the result.
\end{proof}

\subsection{Security of the protocol}
\label{sec:imperfect-security}

We consider a passive computationally unbounded quantum eavesdropper that is
given the complete two-message public transcript $(\publicMask ,z)$ and has quantum
query access to $\randomOracle $.
It succeeds if it outputs the key, which corresponds to Bob's sampled subset $\sharedKey $.
We give the complete transcript to the eavesdropper before it makes any
query.
This gives it at least as much information as receiving the
messages during the protocol.
Throughout this section, the query cost of an eavesdropper is its maximum
number of queries over all transcripts and internal measurement outcomes.
As usual, an algorithm making fewer queries on some branches may be padded
with dummy queries.

We first compare the real experiment with an ideal experiment in
which Bob sends
\[
  z_{\rm ideal}=\sum_{i\in \sharedKey }a_i
\]
using Alice's original labels.
Couple the real and ideal experiments using the
same oracle, masks, labels, and set $\sharedKey $.
The two experiments have the same key $\sharedKey $ and the same oracle.
Their
transcripts agree whenever no uniform block selected by Bob contains an
accidental relation.
Under this coupling, the full data available to Eve
can differ only on that event.
Thus, for every eavesdropper Eve, the success probabilities in
the two experiments differ by at most
\begin{equation}\label{eq:imperfect-real-ideal}
  \eps_{\rm imp}\le\outerArity  \outerBits \securityParam ^{-4}=o(1).
\end{equation}

Fix a public value $z$.  In the ideal experiment, conditioned on this value,
the pair $(\sharedKey ,(a_1,\ldots,a_\securityParam ))$ has distribution $\plantedDist _{\outerArity ,\securityParam }^{z}$.  Indeed,
for every fixed $\sharedKey $, the sum of the initially uniform labels on $\sharedKey $ is
uniform in $\F_2^\outerBits $.  Its probability of being $z$ is $2^{-\outerBits }$, independent
of $\sharedKey $.  Conditioning on $z$ therefore keeps $\sharedKey $ uniform and makes the
labels uniform subject to that sum equation.  By
\Cref{lem:mask-distribution} and~\eqref{eq:single-encoding}, every label bit is
independently encoded by $\nullDist _{\innerArity ,\securityParam }$ when it equals zero and by
$\plantedDist _{\innerArity ,\securityParam }^{0}$ when it equals one.  The masks are independent and uniform in both experiments.
The distinguisher below will sample them as its own randomness, so no
separate composition argument for the public masks is needed.

\begin{theorem}
\label{thm:imperfect-key-recovery}
Let $\innerArity ,\outerArity \ge2$ satisfy
\begin{equation}\label{eq:protocol-conditions}
 \innerArity ^2\securityParam ^{-1/(\innerArity +1)}\le1,
 \qquad
 \outerArity ^2\securityParam ^{-1/(\outerArity +1)}\le1,
\end{equation}
and let $\successProb >0$ be a constant.  Any quantum eavesdropper whose worst-case
query count is $\queryBudget $ and which satisfies
\[
 \Prb[\textnormal{the eavesdropper outputs }\sharedKey ]\ge\successProb
\]
must make
\[
 \queryBudget =\Omega_{\successProb }\!\left(
   \frac{\securityParam ^{\frac \innerArity {\innerArity +1}+\frac\outerArity {\outerArity +1}}}{\sqrt{\innerArity \outerArity }}
 \right)
\]
quantum oracle queries.
\end{theorem}

\begin{proof}
By~\eqref{eq:imperfect-real-ideal}, the same eavesdropper succeeds in the ideal
experiment with probability $\successProb -o(1)$, averaged over the public value $z$.
At least one value $z^*$ has conditional success probability at least
$\successProb -o(1)$.
Use $z^*$ as a fixed public target in the algorithm.
We convert this conditional
algorithm into a distinguisher between the encoded planted and null
distributions of \Cref{thm:single-sum-composition} with target $z^*$.

The null experiment is a comparison distribution used only in the proof:
its original outer labels are independent and uniform, while the public
target is the fixed $z^*$.
It does not need to describe a real protocol execution.
We now describe the distinguisher in three steps.

\emph{Simulating Eve's input.}  Its
encoded input consists of all masked inner strings $Y$.
Independently sample
one uniform public mask string $\publicMask $ for every block, give the eavesdropper
the mask table together with $z^*$, and simulate its random-oracle access using
\[
 \randomOracle =Y+\publicMask .
\]
By \Cref{lem:mask-distribution}, this is exactly the ideal transcript
distribution, and every quantum query to $\randomOracle $ is simulated by one query to
$Y$ followed by coherent addition of the known mask.

\emph{Verifying Eve's candidate.}
Run the eavesdropper on this transcript and let $J$ be its output.
Reject if $J$ is not an $\outerArity $-subset.  Otherwise, read all $\securityParam $ oracle
positions in each of the $\outerArity  \outerBits $ inner blocks indexed by $J$, decode each
block using~\eqref{eq:single-decoder}, and accept exactly when the decoded
labels sum to $z^*$.
This verification procedure makes $\outerArity  \outerBits \securityParam $ queries.

\emph{Comparing the acceptance probabilities.}
On the planted distribution, if the eavesdropper outputs the hidden $\sharedKey $, the
verifier accepts unless one of the selected uniform blocks contains an
accidental $\innerArity $-sum.  Thus its planted acceptance probability satisfies
\[
  \acceptProb _\plantedDist \ge\successProb -o(1)-\outerArity  \outerBits \securityParam ^{-4}=\successProb -o(1).
\]

This subtraction does not assume independence between Eve's success and
the decoding event.
Conditioned on the outer labels and hidden $\sharedKey $, the
selected null blocks are still uniform, so their total error probability
is at most $\outerArity  \outerBits \securityParam ^{-4}$ even for the fixed target $z^*$.

On the encoded null distribution, let the original outer labels be
$a_1,\ldots,a_\securityParam $.  We use a good event for \emph{all} $\securityParam \outerBits $ blocks, rather
than just the blocks in Eve's candidate: Eve may choose that candidate
after querying the oracle.  If no uniform block contains an accidental $\innerArity $-sum,
all decoded labels equal their original values.  Therefore acceptance implies
that the original outer instance has some $\outerArity $-subset summing to $z^*$.  A
union bound gives
\begin{align*}
 \acceptProb _\nullDist
 &\le
   \frac{\binom \securityParam \outerArity }{2^\outerBits }
   +\securityParam \outerBits \frac{\binom \securityParam \innerArity }{2^\innerBits }\\
 &\le \securityParam ^{-3}+\securityParam \outerBits \securityParam ^{-4}
   =o(1).
\end{align*}
By~\eqref{eq:protocol-conditions}, we have
$\outerArity =O(\log \securityParam )$ and $\outerBits =O((\log \securityParam )^2)$, so the last expression tends to
zero.  To see how this gives a useful adversary numerator, write
\begin{align*}
 \outputOverlap (\acceptProb _\plantedDist ,\acceptProb _\nullDist )
 &=\sqrt{\acceptProb _\plantedDist \acceptProb _\nullDist }+\sqrt{(1-\acceptProb _\plantedDist )(1-\acceptProb _\nullDist )}\\
 &\le\sqrt{\acceptProb _\nullDist }+\sqrt{1-\acceptProb _\plantedDist }\\
 &\le\sqrt{\acceptProb _\nullDist }+1-\frac{\acceptProb _\plantedDist }{2}.
\end{align*}
The last inequality uses $\sqrt{1-x}\le1-x/2$ for $x\in[0,1]$.
Therefore
\[
 1-\outputOverlap (\acceptProb _\plantedDist ,\acceptProb _\nullDist )\ge\frac{\acceptProb _\plantedDist }{2}-\sqrt{\acceptProb _\nullDist }
                   \ge\frac\successProb 2-o(1).
\]
For fixed $\successProb >0$, this is at least $\successProb /4$ for all sufficiently large $\securityParam $.
By \Cref{thm:single-sum-composition}, the distinguisher therefore makes
\[
 \Omega_{\successProb }\!\left(
   \frac{\securityParam ^{\frac \innerArity {\innerArity +1}+\frac\outerArity {\outerArity +1}}}{\sqrt{\innerArity \outerArity }}
 \right)
\]
queries.
If the original eavesdropper uses $\queryBudget $ queries, the distinguisher
uses $\queryBudget +\outerArity  \outerBits \securityParam $.  We now check that the verification cost is smaller than the lower bound
by a factor that tends to zero.
Condition~\eqref{eq:protocol-conditions} implies
$\securityParam \ge \innerArity ^{2(\innerArity +1)}$ and $\securityParam \ge\outerArity ^{2(\outerArity +1)}$, and in particular
$\innerArity ,\outerArity =O(\log \securityParam )$.  Therefore $\outerArity  \outerBits \securityParam =O(\securityParam (\log \securityParam )^3)$.  On the other hand,
\[
 \frac{\securityParam ^{\frac \innerArity {\innerArity +1}+\frac\outerArity {\outerArity +1}}}{\sqrt{\innerArity \outerArity }}
 \ge \frac{\securityParam ^{4/3}}{O(\log \securityParam )}.
\]
In particular, the ratio of the verification cost to the lower-bound
scale is
\[
 O\!\left(\frac{\securityParam (\log \securityParam )^3\log \securityParam }{\securityParam ^{4/3}}\right)
 =O\!\left(\frac{(\log \securityParam )^4}{\securityParam ^{1/3}}\right)=o(1).
\]
We can therefore subtract the verification cost and keep the stated
lower bound for $\queryBudget $.
\end{proof}

The theorem also shows when the recovery probability must tend to zero.
Write
\[
 \mathcal L=
 \frac{\securityParam^{\innerArity/(\innerArity+1)+\outerArity/(\outerArity+1)}}
      {\sqrt{\innerArity\outerArity}}.
\]
If a family of eavesdroppers uses $\queryBudget=o(\mathcal L)$ queries,
its recovery probability tends
to zero.  Otherwise some constant $\successProb >0$ would be achieved along an
infinite sequence of input lengths, contradicting the theorem on that
sequence.  This concerns the actual distribution of protocol executions,
with a worst-case bound on each eavesdropper's query count.

%% file: tex/perfect2.tex
\section{Perfect-agreement upgrade}\label{sec:perfect}

We replace every one-block bit encoding by a pair of blocks and add a
final confirmation message.
If decoding fails or Alice cannot identify a
unique subset, both parties output the fixed key $\fallbackKey$.
Thus the protocol always outputs a key and has perfect agreement.
The public transcript reveals when this fallback key is used, but this
event has vanishing probability.
We include these executions in the
security analysis.  The paired transformation preserves the relevant
adversary ratio, so the security exponent is unchanged.

Consider a random oracle
\[
 \widetilde \randomOracle :[\securityParam ]\times[\outerBits ]\times\{L,R\}\times[\securityParam ]
       \longrightarrow\F_2^\innerBits .
\]
Thus there are $2\securityParam \outerBits $ independent length-$\securityParam $ blocks.  Alice prepares each block
with the masking procedure of \Cref{lem:mask-distribution}, applied to the
corresponding restriction of $\widetilde \randomOracle $.

\subsection{Paired encoding and the three-message protocol}

To encode $b\in\{0,1\}$, Alice prepares the left and right masked blocks by
\begin{equation}\label{eq:paired-encoding}
 b=0:\quad (Y^L,Y^R)\leftarrow \plantedDist _{\innerArity ,\securityParam }^{0}\otimes \nullDist _{\innerArity ,\securityParam },
 \qquad
 b=1:\quad (Y^L,Y^R)\leftarrow \nullDist _{\innerArity ,\securityParam }\otimes \plantedDist _{\innerArity ,\securityParam }^{0}.
\end{equation}
Bob decodes the pair by
\begin{equation}\label{eq:paired-decoder}
 \pairDecoder (Y^L,Y^R)=
 \begin{cases}
  0,&\relationTest (Y^L)=1\text{ and }\relationTest (Y^R)=0,\\
  1,&\relationTest (Y^L)=0\text{ and }\relationTest (Y^R)=1,\\
  \bot,&\text{otherwise}.
 \end{cases}
\end{equation}
Because one half is planted, the two relation predicates can never both be
zero.  Thus the decoder either returns the encoded bit or returns $\bot$; it
never returns the wrong bit.  The symbol $\bot$ is an internal decoder
outcome; it leads to the shared fallback key in the protocol below.

The key space is
\[
 \mathcal K=\{\fallbackKey\}\sqcup
             \binom{[\securityParam]}{\outerArity}.
\]
Here $\sqcup$ denotes disjoint union: $\fallbackKey$ is a separate key
value, distinct from every subset key.  We keep $\sharedKey$ for Bob's
sampled subset and write $\agreedKey_A,\agreedKey_B$ for the parties'
actual outputs.  The message $\mathsf{fallback}$ is also a separate
symbol, distinct from every $z\in\F_2^{\outerBits}$.

The protocol is as follows.
\begin{enumerate}[leftmargin=2.2em,label=\textbf{\arabic*.}]
\item Alice samples independent uniform labels
      $a_1,\ldots,a_\securityParam\in\F_2^\outerBits$.
      For every $(i,t)\in[\securityParam]\times[\outerBits]$, she
      encodes $a_i(t)$ by~\eqref{eq:paired-encoding} and sends all masks
      to Bob.
\item Bob samples a uniform
      $\sharedKey\in\binom{[\securityParam]}{\outerArity}$, reads both
      halves of every pair $(i,t)$ with $i\in\sharedKey$, and applies
      \eqref{eq:paired-decoder}.
      If any result is $\bot$, he sends
      $\mathsf{fallback}$.
      Otherwise he obtains labels
      $\widehat a_i\in\F_2^\outerBits$ and sends
      \[
        z=\sum_{i\in\sharedKey}\widehat a_i.
      \]
\item If Bob sent $\mathsf{fallback}$, Alice replies
      $\mathsf{fallback}$ and outputs $\agreedKey_A=\fallbackKey$.
      Otherwise she finds all $\outerArity$-subsets $J$ satisfying
      $\sum_{j\in J}a_j=z$.
      If there is exactly one, she replies
      $\mathsf{accept}$ and outputs $\agreedKey_A=J$.
      If there is no unique solution, she replies $\mathsf{fallback}$
      and outputs $\agreedKey_A=\fallbackKey$.
      Upon receiving Alice's reply, Bob outputs
      \[
       \agreedKey_B=
       \begin{cases}
        \sharedKey,&\text{if the reply is }\mathsf{accept},\\
        \fallbackKey,&\text{if the reply is }\mathsf{fallback}.
       \end{cases}
      \]
\end{enumerate}
Let $\fallbackEvent$ be the event that Alice's final message is
$\mathsf{fallback}$.

\begin{proposition}%
\label[proposition]{prop:perfect-correctness}
For every oracle and every choice of the parties' randomness, both
parties output a key and $\agreedKey_A=\agreedKey_B$.
Write $\agreedKey$ for this common key.  Then
\begin{equation}\label{eq:perfect-fallback-probability}
 \Prb[\agreedKey=\fallbackKey]
 =\Prb[\fallbackEvent]
 \le\eps_{\rm perf}
 :=\outerArity\,\outerBits\,\securityParam^{-4}
       +\securityParam^{-3}.
\end{equation}
Alice makes exactly $\innerArity\securityParam\outerBits$ oracle queries
and Bob makes at most $2\outerArity\,\outerBits\securityParam$
queries.  Thus the total worst-case query complexity is
$O((\innerArity+\outerArity)\outerArity\,\securityParam\log\securityParam)$.
\end{proposition}

\begin{proof}
For every encoded bit, one half of the pair is planted and therefore
has a relation.
If the other half has no relation, Bob decodes the
original bit; if it has a relation, he obtains $\bot$.
Therefore, whenever Bob sends $z$, all his decoded labels equal
Alice's labels and $\sharedKey$ is a solution of Alice's equation.
If Alice replies $\mathsf{accept}$, that solution is unique, so
\[
 \agreedKey_A=J = \sharedKey=\agreedKey_B.
\]
If Alice replies $\mathsf{fallback}$, both parties output
$\fallbackKey$.
This proves perfect agreement for every oracle and
every choice of randomness.

Among the $\outerArity\,\outerBits$ pairs selected by Bob, there is
one uniform half-block per pair.
By~\eqref{eq:false-positive} and a
union bound, the probability that any of these uniform halves contains
an accidental relation is at most
$\outerArity\,\outerBits\securityParam^{-4}$.
Also, the unconditional probability that there is a second outer
subset $J\ne\sharedKey$ with
$\sum_{j\in J}a_j=\sum_{i\in\sharedKey}a_i$ is at most
$\securityParam^{-3}$, by the uniqueness argument in
\Cref{prop:imperfect-correctness}.
If neither event occurs, Bob decodes every selected bit correctly
and Alice finds the unique solution $\sharedKey$.
Thus
$\fallbackEvent$ implies at least one of these events, proving
\eqref{eq:perfect-fallback-probability}.

Alice prepares exactly one planted half per label bit, at a cost of
$\innerArity$ queries.
Bob reads two length-$\securityParam$ halves
per selected bit.
The query bounds follow from
$\outerBits=O(\outerArity\log\securityParam)$.
\end{proof}

\subsection{Security of the protocol}

An eavesdropper receives the complete three-message transcript and
outputs a guess $\widehat{\agreedKey}$ for the common key $\agreedKey$.
Its success probability is
\[
 \Prb[\widehat{\agreedKey}=\agreedKey],
\]
measured over all executions.
On $\fallbackEvent$, the final message
reveals that $\agreedKey=\fallbackKey$, so the eavesdropper can succeed
without making any query.
These executions contribute at most
$\eps_{\rm perf}$ to its overall success probability.
As before, its query cost is the maximum number of quantum queries over
all transcripts and internal outcomes.

Consider an ideal experiment in which Bob sends
\[
 z_{\rm ideal}=\sum_{i\in\sharedKey}a_i,
\]
Alice's last message is always $\mathsf{accept}$, and the target key is
always $\sharedKey$.  Couple it to the real experiment using the same
oracle, masks, labels, and sampled subset $\sharedKey$.
Outside $\fallbackEvent$, the oracle, transcript, and target key are
identical in the two experiments.  Thus any fixed eavesdropper satisfies
\begin{equation}\label{eq:perfect-real-ideal}
 \left|
 \Prb_{\rm real}[\widehat{\agreedKey}=\agreedKey]
 -\Prb_{\rm ideal}[\widehat{\agreedKey}=\sharedKey]
 \right|
 \le\Prb[\fallbackEvent]\le\eps_{\rm perf}.
\end{equation}
In particular, the change of experiment accounts for all successes
obtained by guessing the fallback key.

Conditioned on a fixed public value $z$, the ideal outer instance has
distribution $\plantedDist_{\outerArity,\securityParam}^{z}$, and each
bit is independently encoded according to~\eqref{eq:paired-encoding}.
The masks are independent public randomness and do not affect the
adversary ratio.

\begin{theorem}[Perfect-agreement key-recovery lower bound]
\label{thm:perfect-key-recovery}
Suppose $\innerArity ,\outerArity \ge2$ satisfy~\eqref{eq:protocol-conditions}, and let $\successProb >0$
be a constant.  Any quantum eavesdropper whose worst-case query count is
$\queryBudget$ and whose unconditional key-recovery probability satisfies
\[
 \Prb[\widehat{\agreedKey}=\agreedKey]\ge\successProb
\]
must have
\[
 \queryBudget =\Omega_{\successProb }\!\left(
   \frac{\securityParam ^{\frac \innerArity {\innerArity +1}+\frac\outerArity {\outerArity +1}}}{\sqrt{\innerArity \outerArity }}
 \right)
\]
quantum random-oracle queries.
\end{theorem}

\begin{proof}
By~\eqref{eq:perfect-real-ideal}, the eavesdropper succeeds in the ideal
experiment with probability at least $\successProb-\eps_{\rm perf}$.
Under~\eqref{eq:protocol-conditions}, we have $\eps_{\rm perf}=o(1)$.
Averaging over $z$, fix a value $z^*$ for which the conditional success
probability is at least $\successProb-\eps_{\rm perf}$.
We turn this conditional algorithm into a distinguisher for the paired
encoded distributions in \Cref{thm:paired-sum-composition} with target
$z^*$.

On its encoded input, the distinguisher samples an independent uniform mask for
every half-block, gives the eavesdropper the mask table, $z^*$, and the final
message $\mathsf{accept}$, and simulates the oracle as
$\widetilde \randomOracle =Y+\publicMask $.  By \Cref{lem:mask-distribution}, this is exactly the
ideal conditional transcript, and each oracle query costs one query to $Y$.
Let $J$ be the eavesdropper's output.  Reject unless $J$ is an
$\outerArity$-subset; in particular, reject the output $\fallbackKey$.
Otherwise read both halves of all $\outerArity  \outerBits $ pairs indexed by $J$, reject if any
paired decoder returns $\bot$, and accept exactly if the decoded labels sum
to $z^*$.
This verification procedure costs $2\outerArity  \outerBits \securityParam $ queries.

On the planted distribution, when the eavesdropper outputs the hidden $\sharedKey $,
verification fails only if a selected uniform half-block contains an
accidental relation.  Therefore
\[
 \acceptProb_\plantedDist
 \ge\successProb-\eps_{\rm perf}
      -\outerArity\,\outerBits\securityParam^{-4}
 =\successProb-o(1).
\]
On the null distribution, a paired decoding that does not return $\bot$ always equals its
original outer bit.  Acceptance therefore implies that the original outer
instance has an $\outerArity $-subset summing to $z^*$, and thus
\[
 \acceptProb _\nullDist \le\frac{\binom \securityParam \outerArity }{2^\outerBits }\le \securityParam ^{-3}=o(1).
\]
The same elementary estimate used in the imperfect case gives
\[
 1-\outputOverlap (\acceptProb _\plantedDist ,\acceptProb _\nullDist )\ge\frac{\acceptProb _\plantedDist }{2}-\sqrt{\acceptProb _\nullDist }
                  \ge\frac\successProb 2-o(1).
\]
The paired composition theorem therefore gives
\[
 \queryBudget +2\outerArity  \outerBits \securityParam
 =\Omega_\successProb \!\left(
   \frac{\securityParam ^{\innerArity /(\innerArity +1)+\outerArity /(\outerArity +1)}}{\sqrt{\innerArity \outerArity }}\right).
\]
The verification cost is twice the cost in the imperfect case.  Its ratio
to the lower-bound scale is still
$O((\log \securityParam )^4/\securityParam ^{1/3})=o(1)$, so we can subtract it.  This proves the
claimed bound on $\queryBudget $.
\end{proof}

%% file: tex/near-quadratic2.tex
\section{Parameter choices}
\label{sec:parameters}

\subsection{Near-quadratic separation}
\label{sec:quadratic-sep}

\begin{theorem}
  \label{thm:uniform}
There are two families of classical random oracle protocols:
\begin{enumerate}[leftmargin=2.2em,label=\textnormal{(\roman*)}]
\item a two-message protocol in which both parties output the same key with
      probability $1-O(\securityParam ^{-3})$; and
\item a three-message protocol that always outputs the same key at both
      parties, with $\Prb[\agreedKey=\fallbackKey]=O(\securityParam^{-3})$.
\end{enumerate}
Both have honest query complexity
\[
 O\!\left(
   \frac{\securityParam (\log \securityParam )^3}{(\log\log \securityParam )^2}
 \right)=\widetilde O(\securityParam ),
\]
and every quantum eavesdropper recovering Bob's key with constant
probability must make at least
\begin{equation}\label{eq:uniform-main-bound}
 \Omega_{\successProb }\!\left(
   \frac{\securityParam ^2\log\log \securityParam }{(\log \securityParam )^{5}}
 \right)=\widetilde\Omega_{\successProb }(\securityParam ^2)
\end{equation}
queries.
In terms of the honest query budget $\honestBudget $, the lower bound is
$\widetilde\Omega_{\successProb }(\honestBudget ^2)$.
\end{theorem}

\begin{proof}
For all sufficiently large $\securityParam $, set
\[
 \sumArity =\left\lfloor\frac{\log \securityParam }{2\log\log \securityParam }\right\rfloor
\]
and take $\innerArity =\outerArity =\sumArity $.  Write $\lambda=\log \securityParam $ and
$h=\log\log \securityParam $.  The floor in the definition of $\sumArity $ gives
$\sumArity +1=\lambda/(2h)+\theta$ for some $0<\theta\le1$.  Therefore
\begin{align*}
  \frac{\lambda}{\sumArity +1}
  &= \frac{\lambda}{\lambda/(2h) + \theta}\\
 &=\frac{2h}{1+2h\theta/\lambda}\\
 &=2h+O(h^2/\lambda).
\end{align*}
Since $h^2/\lambda\to0$, this proves
\begin{equation}\label{eq:lambda-over-k}
 \frac{\lambda}{\sumArity +1}=2h+o(1).
\end{equation}
Therefore,
\[
 \sumArity ^2\securityParam ^{-1/(\sumArity +1)}
 \le \frac{\lambda^2}{4h^2}
      \exp\!\left(-\frac{\lambda}{\sumArity +1}\right)
 =\frac{e^{o(1)}}{4h^2}<1.
\]
Thus the assumptions of both~\Cref{thm:imperfect-key-recovery} and~\Cref{thm:perfect-key-recovery} hold.

Each theorem gives the lower bound
\begin{align*}
 \Omega_\successProb \!\left(\frac{\securityParam ^{2\sumArity /(\sumArity +1)}}{\sumArity }\right)
 &=\Omega_\successProb \!\left(
     \frac{\securityParam ^2}{\sumArity }
     \exp\!\left(-\frac{2\lambda}{\sumArity +1}\right)
   \right)\\
 &=\Omega_\successProb \!\left(
     \frac{\securityParam ^2h}{\lambda^{5}}
   \right),
\end{align*}
where for the last equality, we use the following substitutions
\[
 \exp\!\left(-\frac{2\lambda}{\sumArity +1}\right)
 =e^{-4h+o(1)}=\lambda^{-4}e^{o(1)},
 \qquad
 \frac1\sumArity =\frac{2h}{\lambda}(1+o(1)).
\]
Multiplying gives $(2 + o(1))\securityParam ^2h/\lambda^{5}$, which proves~\eqref{eq:uniform-main-bound}.

By~\Cref{prop:imperfect-correctness,prop:perfect-correctness}, the honest query
complexity in either case is
\[
 O(\sumArity \securityParam \outerBits )=O(\securityParam \sumArity ^2\log \securityParam )
 =O\!\left(\frac{\securityParam (\log \securityParam )^3}{(\log\log \securityParam )^2}\right).
\]
The imperfect protocol's agreement error and the perfect protocol's
probability of outputting $\fallbackKey$ are each at most
\[
 \sumArity \outerBits \securityParam ^{-4}+\securityParam ^{-3}=O(\securityParam ^{-3}),
\]
because $\sumArity \outerBits =O((\log \securityParam )^3/(\log\log \securityParam )^2)=o(\securityParam )$.
For these implementations, the honest budget satisfies
$\honestBudget =\Theta(\securityParam \sumArity ^2\log \securityParam )=\securityParam \,\operatorname{polylog}\securityParam $ and
$\log \honestBudget =\Theta(\log \securityParam )$.
Replacing $\securityParam $ by $\honestBudget $ therefore changes only
polylogarithmic factors, so
$\widetilde\Omega(\securityParam ^2)=\widetilde\Omega(\honestBudget ^2)$.
\end{proof}

\begin{remark}
For every fixed $\eps>0$, one may instead take a sufficiently large constant
$\sumArity $ and recover the simpler lower bound $\Omega_\eps(\securityParam ^{2-\eps})$ with
$O_\eps(\securityParam \log \securityParam )$ honest queries.
\Cref{thm:uniform} is slightly stronger because it
uses one parameter rule and gives the quadratic exponent up to
polylogarithmic factors.
\end{remark}

\subsection{Efficient implementation with public targets}
\label{sec:public-targets}

\begin{theorem}
\label{thm:efficient-honest-parties}
For every integer $\securityParam\ge64$, write
\( \outerBits=\lceil5\log_2\securityParam\rceil, \innerBits=\lceil6\log_2\securityParam\rceil. \)
There is a two-message classical random oracle protocol with agreement
probability at least $1-\outerBits/\securityParam^4-1/(2\securityParam^3)$, where
each honest party makes at most \(2\securityParam\outerBits\) oracle queries.
Their combined running time is \(O(\securityParam(\log\securityParam)^2)\).
The communication is at most \(\securityParam\outerBits\innerBits+\outerBits\) bits, and
for every constant $\successProb>0$, any passive quantum eavesdropper
recovering Bob's key with probability at least $\successProb$
must make
\(
 \Omega_{\successProb}(\securityParam^{4/3})
\)
queries in the worst case.
\end{theorem}

\begin{proof}
Take $\innerArity=\outerArity=2$ and use $\impProtocol_{2,2}$,
replacing the mask for each block by one public target.

\emph{Encoding and conditional distributions.}
Each block uses disjoint oracle addresses, with values
$Y_j=\randomOracle(j)\in\F_2^{\innerBits}$ for
$j\in[\securityParam]$.
In this variant, no masks are used.
To encode a bit, Alice publishes
\[
 \begin{array}{ll}
 \text{zero:}
   &\blockTarget\leftarrow\Unif(\F_2^{\innerBits}),\\
 \text{one:}
   &S\leftarrow\Unif\!\left(\binom{[\securityParam]}2\right),
     \qquad\blockTarget=\displaystyle\sum_{j\in S}Y_j.
 \end{array}
\]
Encoding zero makes no queries and encoding one makes two queries.
Bob reads the block and decodes one exactly when two distinct
positions sum to $\blockTarget$.
All other protocol steps are unchanged.

For each fixed $S$, its oracle sum is uniform.  Thus the target is
uniform in both modes, and for every $u$,
\[
 \begin{array}{ll}
 \text{zero:}
   &\mathcal L(Y\mid\blockTarget=u)=\nullDist_{2,\securityParam},\\
 \text{one:}
   &\mathcal L(S,Y\mid\blockTarget=u)
       =\plantedDist_{2,\securityParam}^{u}.
 \end{array}
\]
Indeed, conditioning on a planted target $u$ leaves $S$ uniform.
Given $S$ and this target, $Y$ is uniform subject to
$\sum_{j\in S}Y_j=u$.
Since blocks use independent randomness and disjoint oracle addresses,
the public targets are mutually independent and uniform even when all
encoded labels are fixed. Their joint distribution therefore does not
depend on the encoded labels.
Conditioned on the labels and targets, the blocks remain independent,
with the distributions specified above.

\emph{Agreement and security.}
For every fixed target, a null block contains a matching pair with
probability at most $\binom{\securityParam}2\,2^{-\innerBits}$.
A planted block always contains one.
The correctness argument of~\Cref{prop:imperfect-correctness} therefore
applies without change and gives agreement error at most
\[
 \eps_{\rm eff}
 :=\binom{\securityParam}2
       \left(2\outerBits\,2^{-\innerBits}+2^{-\outerBits}\right)
 \le\frac{\outerBits}{\securityParam^4}
       +\frac{1}{2\securityParam^3}.
\]

The security proof also follows the earlier argument.
After conditioning on all public targets, each inner block has its own
fixed sum target.  The norm, exact alignment, and query-filter bounds
of~\Cref{thm:exact-norm,lem:uniform-filter} hold for every such target.
Moreover,~\Cref{lem:completed-upper} allows a different inner operator
in each coordinate, and the distribution-vector calculation in~\Cref{lem:composition} applies coordinate by coordinate.
Since $\securityParam\ge64$, the assumptions of these results hold.
Thus the bound of~\Cref{thm:single-sum-composition} holds for these
conditional distributions.

We can now repeat the real-to-ideal comparison and candidate
verification from~\Cref{thm:imperfect-key-recovery}, averaging over both
the public-target vector and Bob's ideal message.
The decoding-error bound is unchanged for every target, so the same
planted and null acceptance bounds apply, including the union bound
over all blocks in the null experiment.
Verification uses $2\outerBits\securityParam$ queries.
Consequently, an eavesdropper with recovery probability at least
$\successProb$ and worst-case query count $\queryBudget$ satisfies
\[
 \queryBudget+2\outerBits\securityParam
   =\Omega_{\successProb}(\securityParam^{4/3}).
\]
Since $\outerBits\securityParam=O(\securityParam\log\securityParam)
=o(\securityParam^{4/3})$, this proves the security claim.

\emph{Honest costs.}
Alice makes at most $2\securityParam\outerBits$ queries to prepare the encodings.
Bob reads
$2\outerBits$ blocks of $\securityParam$ entries each.
Alice sends $\securityParam\outerBits$ targets of $\innerBits$ bits, and Bob sends one $\outerBits$-bit sum.
Thus the communication is \(\securityParam\outerBits\innerBits + \outerBits\).

To decode a block, sort its values and search for the complement $u+y$
of each value $y$.
Keep multiplicities to require distinct positions when $u=0$.
Alice uses the same procedure on her outer labels to find the key and
test uniqueness: count unordered matching pairs, stopping at two.
For distinct complementary value classes of sizes $c,c'$, the number
of pairs is $cc'$; for a class paired with itself, it is $\binom c2$.
Thus no list of all matching pairs is needed.
Sorting, grouping equal values, and binary-searching their complements
take $O(\securityParam\log\securityParam)$ word operations per search.
There are $2\outerBits$ inner searches and one outer search.
Including preparation and communication, the total time is
$O(\securityParam\outerBits\log\securityParam)
=O(\securityParam(\log\securityParam)^2)$ word operations.
\end{proof}

\begin{remark}
A similar result can be obtained with perfect agreement and three messages, by using the paired
encoding and final confirmation from
\Cref{sec:perfect}, with a separate public target for each half-block.
\end{remark}